\documentclass[12pt,a4paper]{article}

\usepackage[british]{babel}
\usepackage[a4paper,top=2cm,bottom=2cm,left=2.5cm,right=2.5cm,marginparwidth=1.75cm]{geometry}
\usepackage{tikz}
\usetikzlibrary{decorations.pathreplacing}
\usepackage[braket, qm]{qcircuit}      % For quantum circuits
\usepackage{graphicx}   

\usepackage{cite}
\usepackage{amsmath}
\usepackage{graphicx}
\usepackage[colorlinks=true, allcolors=blue]{hyperref}
\usepackage{hyperref}
\usepackage[title]{appendix}
\usepackage{mathrsfs}
\usepackage{amsfonts}
\usepackage{booktabs} % For \toprule, \midrule, \botrule
\usepackage{caption}  % For \caption
\usepackage{threeparttable} % For table footnotes
\usepackage{algorithm}
\usepackage{algorithmicx}
\usepackage{algpseudocode}
\usepackage{listings}
\usepackage{enumitem}
\usepackage{chngcntr}
\usepackage{booktabs}
\usepackage{lipsum}
\usepackage{subcaption}
\usepackage{authblk}
\usepackage[T1]{fontenc}    % Font encoding
\usepackage{csquotes}       % Include csquotes
\usepackage{diagbox}

\usepackage{setspace}
\usepackage{titlesec}
\titleformat{\section} % Redefine section numbering format
  {\normalfont\Large\bfseries}{\thesection.}{1em}{}
  
\usepackage{lineno} % Add the lineno package

\rightlinenumbers % Right-align line numbers

\usepackage{float}   % for customizing caption position
\usepackage{caption} % for customizing caption format


\title{Discrete-Time Quantum Random Walk for Epidemiological Modeling}

\author[1]{Sayan Manna}
\author[1]{Nikhil Kowshik}
\author[2]{Sudebkumar Prasant Pal}

\affil[1]{Department of Metallurgical \& Materials Engineering, Indian Institute of Technology, Kharagpur, India}
\affil[2]{Department of Computer Science \& Engineering, Indian Institute of Technology, Kharagpur, India}

\begin{document}
\date{}
\maketitle

\begin{abstract}
We introduce a discrete-time quantum random walk (QRW) framework for spatial epidemic modelling on a two-dimensional square lattice and compare its dynamics to classical random-walk SIR models. In our model, each infected site spawns a quantum walker whose coherent evolution (controlled by an amplitude-splitting coin and conditional shifts) can infect visited susceptible sites with probability $p$ and persists for a lifetime of $\tau$ steps. We perform extensive quantum simulations on finite lattices and compute the basic reproduction number $R_0$ across a broad grid of $(p,\tau)$ values. Results show that QRW dynamics interpolate between diffusive and super-diffusive regimes: at low $p$ the QRW reproduces classical-like $R_0$, while at higher $p$ and $\tau$ ballistic propagation and interference produce markedly larger $R_0$ and non-Gaussian spatial profiles. We compare the QRW $R_0$ range to empirical estimates from historical outbreaks and discuss parameter regimes where QRW offers a closer qualitative match than classical diffusion. We conclude that QRWs provide a flexible, conceptually novel toy model for exploring rapid or heavy-tailed epidemic spread.
\end{abstract}

\section{Introduction}
% The study of epidemic spread through classical random walk models has yielded critical insights into transmission dynamics, reproductive numbers ($R_0$) between localized and widespread outbreaks \cite{chu2021random}, \cite{de_Oliveira_2022}. However, classical models inherently assume diffusive propagation ($\sigma^2 \sim t$), limiting their ability to capture phenomena such as rapid, long-range transmission or interference-driven suppression of outbreaks. 
% Quantum random walks (QRWs) \cite{kempe2003quantum}, with their inherent superposition, entanglement, and ballistic spreading ($\sigma^2 \sim t^2$), offer a fundamentally distinct framework to model disease dynamics. 
% While classical epidemiological models rely stochastic diffusion, QRWs inherently encode non-local correlations and parallel transmission pathways. This work explores the application of quantum walk theory to epidemiology, offering:
% \begin{itemize}
%     \item A framework to model interference-driven herd immunity effects.
%     \item Predictions for critical thresholds where quantum coherence dominates over classical diffusion.
%     \item Insights into optimizing containment strategies by using quantum-inspired suppression mechanisms (e.g., engineered decoherence).
% \end{itemize}
% By contrasting QRW dynamics with classical results, this study aims to identify regimes where quantum effects significantly alter outbreak trajectories---a critical step toward developing hybrid models for next-generation epidemic forecasting.
The study of epidemic spread through classical random walk models has yielded critical insights into transmission dynamics and reproductive numbers ($R_0$) between localized and widespread outbreaks \cite{chu2021random, de_Oliveira_2022}. However, classical models inherently assume diffusive propagation ($\sigma^2 \sim t$), limiting their ability to capture phenomena such as rapid, long-range transmission or interference-driven suppression of outbreaks. Quantum random walks (QRWs) \cite{kempe2003quantum}, with their inherent superposition, entanglement, and ballistic spreading ($\sigma^2 \sim t^2$), offer a fundamentally distinct framework to model disease dynamics. While classical epidemiological models rely on stochastic diffusion, QRWs inherently encode non-local correlations and parallel transmission pathways. In this work, we explore the application of quantum walk theory to epidemiology by developing a framework to model interference-driven herd immunity effects, predicting critical thresholds where quantum coherence dominates over classical diffusion, and offering insights into optimizing containment strategies through quantum-inspired suppression mechanisms such as engineered decoherence. By contrasting QRW dynamics with classical results, this study aims to identify regimes where quantum effects significantly alter outbreak trajectories---a critical step toward developing hybrid models for next-generation epidemic forecasting.

%%%%%%%%%%%%%%%%%%%%%%%%%%%%%%%%%%%%%%%%%%%%%%%%%%%%%%%
% \section{Simulation on Classical Random Walk}
% \begin{figure}[h]
%     \centering
%     \includegraphics[width=0.5\textwidth]{Classicalexp1.png}

% \includegraphics[width=0.7\textwidth]{Classicalexp2.png}
%     \caption{ \(p = 0.5\) and \(\tau = 2\)}
%     \label{fig:cat1}
% \end{figure}

% We use an infinite 2D square lattice to model the population's geographic distribution. Each site may be in one of three states at any given moment. Despite the basic differences between the model provided here and a SIR model, the same nomenclature is used: each site in the lattice is either susceptible, infected, or eliminated. Except for the origin, which has a single infectious agent, all sites are initially set to be susceptible. A fresh random walker is created by an infected website and begins to wander over the lattice. The walker infects a susceptible site with probability \(p\) if it lands there (sites are shown in the diagram as square cells). The infected site starts a new walker and itself becomes removed (no longer susceptible to infection). The walker heals and stops infecting other locations after taking \(\tau\) steps. In the same way as in well-mixed SIR models, the term "removed" is used to describe areas that are no longer susceptible to infection and are not physically removed from the lattice.

%%%%%%%%%%%%%%%%%%%%%%%%%%%%%%%%%%%%%%%%%%%%%%%%%%%%%%%%%%%%

\section{Discrete Quantum Random Walk}
A quantum random walk (QRW) is the quantum analogue of a classical random walk, where unitary evolution governed by superposition and interference leads to ballistic rather than diffusive spreading \cite{kempe2003quantum,venegas2012quantum}. 

In the discrete-time formulation, the walker evolves on a composite Hilbert space $\mathcal{H} = \mathcal{H}_C \otimes \mathcal{H}_P,
$ with a coin register $\mathcal{H}_C$ (e.g., for 1D, $\mathcal{H}_C = \text{span}\{ \vert \uparrow \rangle, \vert \downarrow \rangle \}
$) controlling movement on the position space $\mathcal{H}_P$ (e.g., for 1D, $\mathcal{H}_P = \text{span}\{ \vert i \rangle : i \in \mathbb{Z} \}$). Each step consists of applying a unitary coin operator $(C)$ (e.g., Hadamard (H)) to create a superposition of directions, followed by a conditional shift $(S)$ that updates the position accordingly. The time evolution operator is
\[
U = S \cdot (C \otimes I),
\]
where, $I$ is identity operator. So the state evolves as $\vert \Psi(t+1) \rangle = U \, \vert \Psi(t) \rangle$. For example, shift operator $S$ in one dimension is given by
\[
S = \sum_{i \in \mathbb{Z}} \Big( \vert \uparrow \rangle \langle \uparrow \vert \otimes \vert i+1 \rangle \langle i \vert \;+\; \vert \downarrow \rangle \langle \downarrow \vert \otimes \vert i-1 \rangle \langle i \vert \Big).
\]

This mechanism generates interference patterns and ballistic spread ($\sigma^2 \sim t^2$), which fundamentally distinguishes QRWs from their classical counterparts and motivates their use in modeling epidemic propagation.

\section{Two-Dimensional Quantum Random Walk}

A two-dimensional quantum random walk (QRW) \cite{mackay2002quantum} generalizes the one-dimensional case to motion on a square lattice. 
The walker is associated with a position on the lattice, labeled by coordinates $(x,y) \in \mathbb{Z}^2$, and its evolution proceeds in discrete time steps according to unitary dynamics. 
At each step, the quantum state of the walker spreads simultaneously along different lattice directions due to superposition, and the resulting probability distribution arises from interference between multiple paths.

Unlike the classical two-dimensional random walk, where the probability distribution remains Gaussian and spreads diffusively, the quantum walk on a square lattice exhibits ballistic propagation and highly non-classical spatial patterns \cite{tregenna2003controlling}.
These features make the two-dimensional QRW a rich model for studying spreading processes on networks and lattices, with applications ranging from quantum transport to epidemic modeling on structured populations.
\subsection{Quantum Circuit Simulation of 2D Quantum Random Walk}

To simulate a two-dimensional quantum random walk on a square lattice, we extend the quantum circuit model of the one-dimensional case (given in Appendix~\ref{appendix:1d}) by incorporating movement along both the $x$- and $y$-directions. 
This requires two coin qubits, which together define four possible coin states: $\{ \lvert 00 \rangle, \lvert 01 \rangle, \lvert 10 \rangle, \lvert 11 \rangle \}$

\begin{figure}[H]
    \centering
    \includegraphics[width=0.6\textwidth]{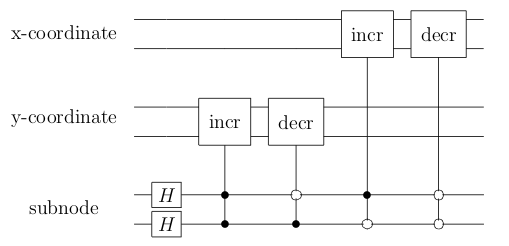}
    \caption{Quantum circuit to implement 2D quantum random walk \cite{douglas2009efficient}}
    \label{fig:2D-QRW-circuit}
\end{figure}

Each coin state determines a unique direction of motion on the lattice:
\[
\begin{aligned}
\lvert 00 \rangle &\;\; \longrightarrow \;\; \text{step in negative $x$ direction}, \\
\lvert 10 \rangle &\;\; \longrightarrow \;\; \text{step in positive $x$ direction}, \\
\lvert 01 \rangle &\;\; \longrightarrow \;\; \text{step in negative $y$ direction}, \\
\lvert 11 \rangle &\;\; \longrightarrow \;\; \text{step in positive $y$ direction}.
\end{aligned}
\]

The circuit implementation \cite{douglas2009efficient} is illustrated in Fig.~\ref{fig:2D-QRW-circuit}. 
Here, the two Hadamard gates prepare the coin qubits in superposition, and the subsequent controlled operations conditionally shift the position registers along the $x$- or $y$-axes. 
The operators labeled \texttt{incr} and \texttt{decr} correspond to unitary translations: \texttt{incr} moves the walker one step in the positive direction along the corresponding axis, while \texttt{decr} moves it in the negative direction. 

By iterating this unitary evolution, the walker spreads coherently over the two-dimensional lattice, and the resulting probability distribution arises from quantum interference between the multiple possible paths. Also, 3-Dimensional quantum walk simulation has been shown in Appendix~\ref{appendix:3d}.

\section{Methodology}
\subsection{Disease Spread Using Quantum Walks}

To model epidemic dynamics, we adopt the classical susceptible--infected--removed (SIR) framework, reformulated in terms of spatial processes on a two-dimensional square lattice \cite{durrett1995spatial}. In this representation, each lattice site corresponds to a geographic location that may exist in one of three possible states at any given time. A site may be \textit{susceptible}, meaning it is capable of being infected by contact with an infectious agent. Alternatively, it may be \textit{infected}, indicating that it currently hosts the agent and has the ability to transmit the infection. Finally, a site can be \textit{removed}, signifying that it has already been infected in the past and is no longer susceptible to further infection.  
Initially, all lattice sites are susceptible except for the origin, which hosts a single infectious agent. In analogy to the SIR model, infection spreads through the random motion of individuals \cite{de_Oliveira_2022}. In the classical random walk formulation, an infected site spawns a walker that traverses the lattice, infecting a susceptible site with probability $p$ upon contact. Once infected, a site immediately transitions to the removed state and simultaneously launches a new walker. Each walker remains infectious for $\tau$ steps, after which it is deactivated. The term ``removed'' refers to immunity or lack of further participation in the epidemic process, rather than physical removal from the lattice. 

\subsubsection{Quantum Walk-Based Epidemic Dynamics}

In the quantum formulation, the walker is replaced by a quantum random walker (QRW) on a two-dimensional square lattice. The walker evolves coherently across both the $x$- and $y$-directions under unitary dynamics, exploiting quantum superposition and interference. The infection rules are defined analogously: at each time step, with probability $p$, a site visited by the quantum walker becomes infected. This infected site then generates a new quantum walker, which also persists for $\tau$ steps before being removed. In this way, infection branches into multiple coherent paths, leading to a fundamentally different spreading profile compared to the classical case.  

% \subsubsection{Percolation-Theoretic Interpretation}

% The epidemic model described above is inspired by percolation theory \cite{grassberger1983critical}. Not all lattice sites need to be populated with susceptible individuals; instead, we assume $N$ agents are randomly distributed across an $L \times L$ lattice, with $N-1$ agents initially susceptible and one agent infected. The central question is whether the infection percolates through the population, which depends critically on the infection probability $p$ and the population density $N/L^2$.  

% We define a critical threshold $p_c$, analogous to percolation thresholds in statistical physics, beyond which the epidemic spreads across a macroscopic fraction of the lattice. Identifying this threshold and its dependence on system size constitutes a central goal of the analysis. In addition, we compare the universality class of the QRW-based epidemic model to that of the classical random walk, seeking differences in scaling exponents.  
\subsubsection{Reproduction number $R_0$ for the QRW model}
We estimate the basic reproduction number \(R_0\) for the QRW-driven process directly from numerical experiments. Here \(R_0\) is defined  as the expected number of \emph{direct} secondary infections produced by a single infected agent introduced into an  fully susceptible population. For each parameter pair \((p,\tau)\) we perform many independent realizations with a single infected agent initially placed at the origin and all other agents susceptible. The QRW dynamics are run until no active walkers remain; in each realization we record the number of distinct susceptible sites that were infected directly by the initial infected agent during its infectious lifetime (first-generation infections). Averaging this first-generation count across realizations yields the simulation estimate of \(R_0(p,\tau)\).
An identical simulation protocol is used for the classical random-walk benchmark (classical walkers with the same \(\tau\) and infection probability \(p\)) \cite{chu2021random}, allowing a direct comparison between quantum-coherent and classical spreading mechanisms.

\subsubsection{Classical vs. Quantum Spreading Profiles}
A key theoretical distinction between classical and quantum walks lies in the spatial probability profile. In the classical case, the probability distribution after many steps converges to an approximately Gaussian form, producing a “hill-shaped” spreading pattern centered at the origin \cite{feller1968introduction}. In contrast, the quantum random walk exhibits interference effects that suppress return probabilities at the origin, leading to a ``bowl-shaped'' distribution in two dimensions \cite{venegas2012quantum}. This alters the disease spread profile in case quantum case from classical case.

% This altered geometry shifts the effective percolation threshold, suggesting that the epidemic transition occurs at different values of $N$ and $p_c$ in the quantum case compared to the classical case.
\begin{figure}[H]
\centering
\includegraphics[width=0.6\linewidth]{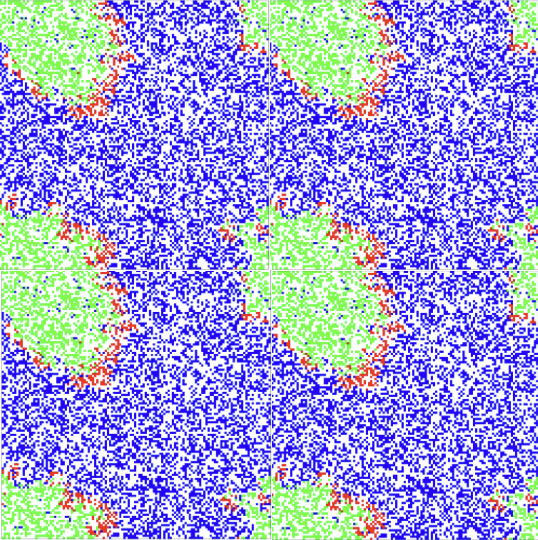}
\caption{\label{fig:result_1}The Cluster of the \(194^{th}\) iteration in a $256\times256$ grid. The skewed and non circular nature of the geometry of the cluster formed using the Quantum Random Walk method shows the stark difference between Classical and Quantum Epidemiology}
\end{figure}

In simulation figures of disease spread, we have marked the susceptible people in blue, the unoccupied sites in white, the active infected agents in red, the recovered sites in green. Also, yellow mark denotes the visited areas by the infected agents. 
Figure~\ref{fig:result_1} displays a representative realization of the quantum-random-walk (QRW) epidemic simulation on a $256\times256$ grid (194\textsuperscript{th} iteration). The outbreak generates compact clusters of infected sites, but their shape is far from uniform. The outer rim of active infections (red) appears stronger in certain directions while being weaker in others, giving the cluster a distinctly skewed pattern. This directional imbalance reflects the non-Gaussian, interference-driven nature of quantum walk propagation.
 By contrast, a classical random walk on the same lattice would produce an essentially isotropic, approximately Gaussian visitation profile with no preferred spreading direction. The directional bias and irregular cluster geometry observed here are therefore direct visual signatures of quantum-coherent motion (superposition and interference) altering the spatial statistics of outbreak propagation.

\subsection{Simulation of Disease spread using QRW}
Simulations are performed on an $L\times L$ square lattice in which $N$ agents are placed uniformly at random on lattice sites.  The initial condition contains $N-1$ susceptible agents and a single infected agent at the origin. An infected site launches a walker which, during each of its at most $\tau$ time steps, attempts to infect any visited susceptible site with probability $p$. Walkers are deactivated after $\tau$ steps or when no further propagation is possible. Simulations are run until no active walkers remain.

In the quantum random walk (QRW) epidemic model, each actively infected agent (shown in red) is constrained to move only along the four cardinal directions: up, down, left, and right. The susceptible agents (shown in blue) are randomly distributed across the lattice. After completing its maximum allowed lifetime $\tau_{\max}$, the walker recovers and the site where it originated is marked as recovered (green). The process begins with a single infected agent at a chosen initial position (e.g., $[1,1]$), together with $N$ susceptible agents distributed across the grid. The epidemic formally terminates once no infected agents remain active. 

The movement of each infected walker is determined by two quantum coins: one governs motion along the horizontal axis, while the other governs motion along the vertical axis. At the $i^{\text{th}}$ iteration, the Qiskit Aer simulator is employed to generate a frequency histogram corresponding to the four possible movement directions. This is obtained using 1024 measurement shots, with the COBYLA optimizer guiding the variational circuit. The walker’s next step is then sampled from this histogram, such that directions with higher frequencies are more likely to be chosen, while those with lower frequencies remain possible, although with reduced probability. This procedure ensures that interference effects inherent to quantum walks are faithfully captured in the infection dynamics. 

During the course of the simulation, all sites visited by any infected walker are classified as visited (yellow). The total number of such distinct visited sites in a single run defines the cluster size, denoted by $M$. To examine the dependence of outbreak size on population density, the simulation is repeated 100 times for each value of $N$, and the average cluster size $\langle M \rangle$ is recorded. 

\begin{figure}[H]
    \centering
    \begin{minipage}{0.30\textwidth}
        \centering
        \includegraphics[width=\textwidth]{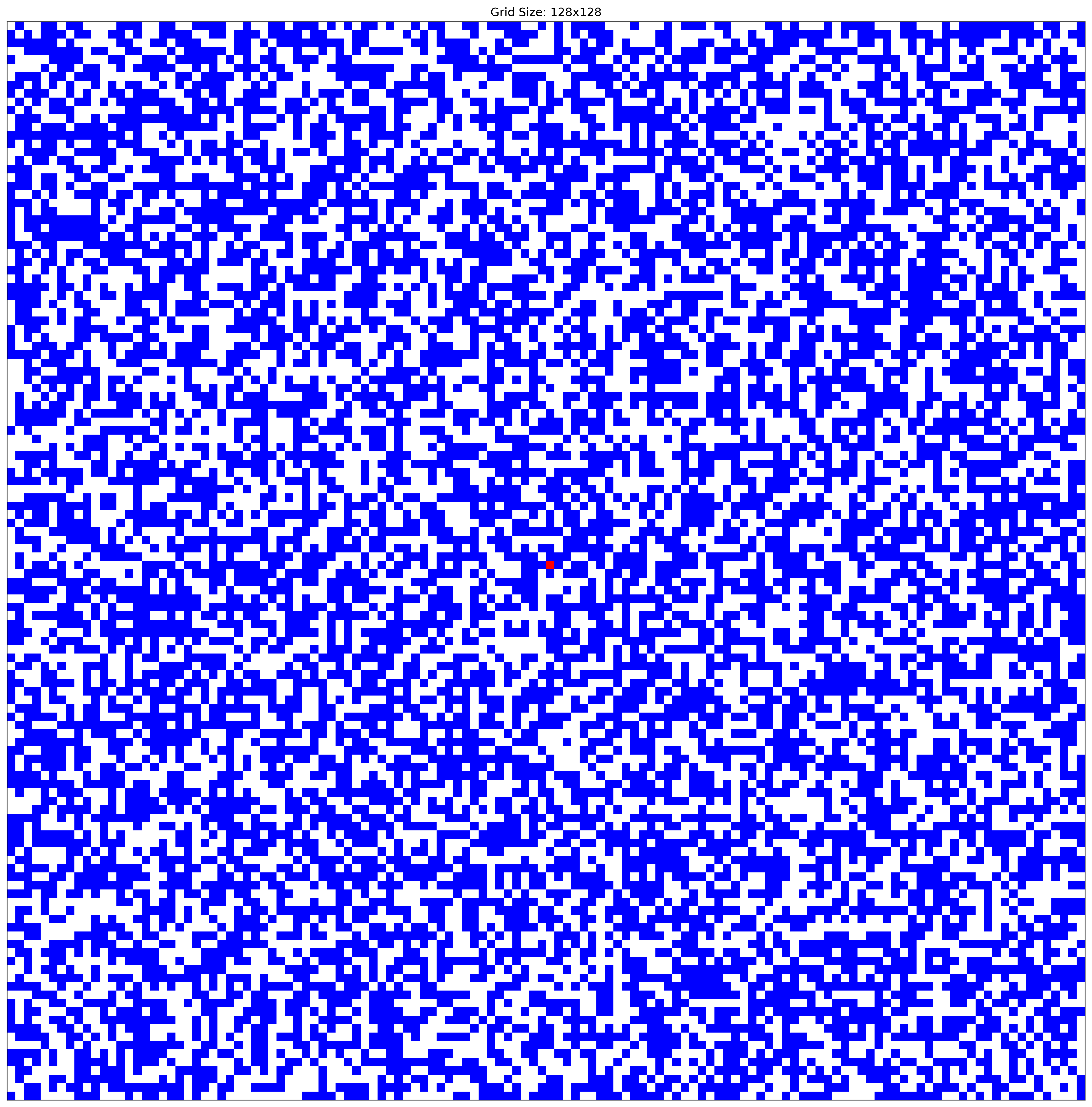}
        \caption{Step = 0}
        \label{fig:step0}
    \end{minipage}%
    \begin{minipage}{0.30\textwidth}
        \centering
        \includegraphics[width=\textwidth]{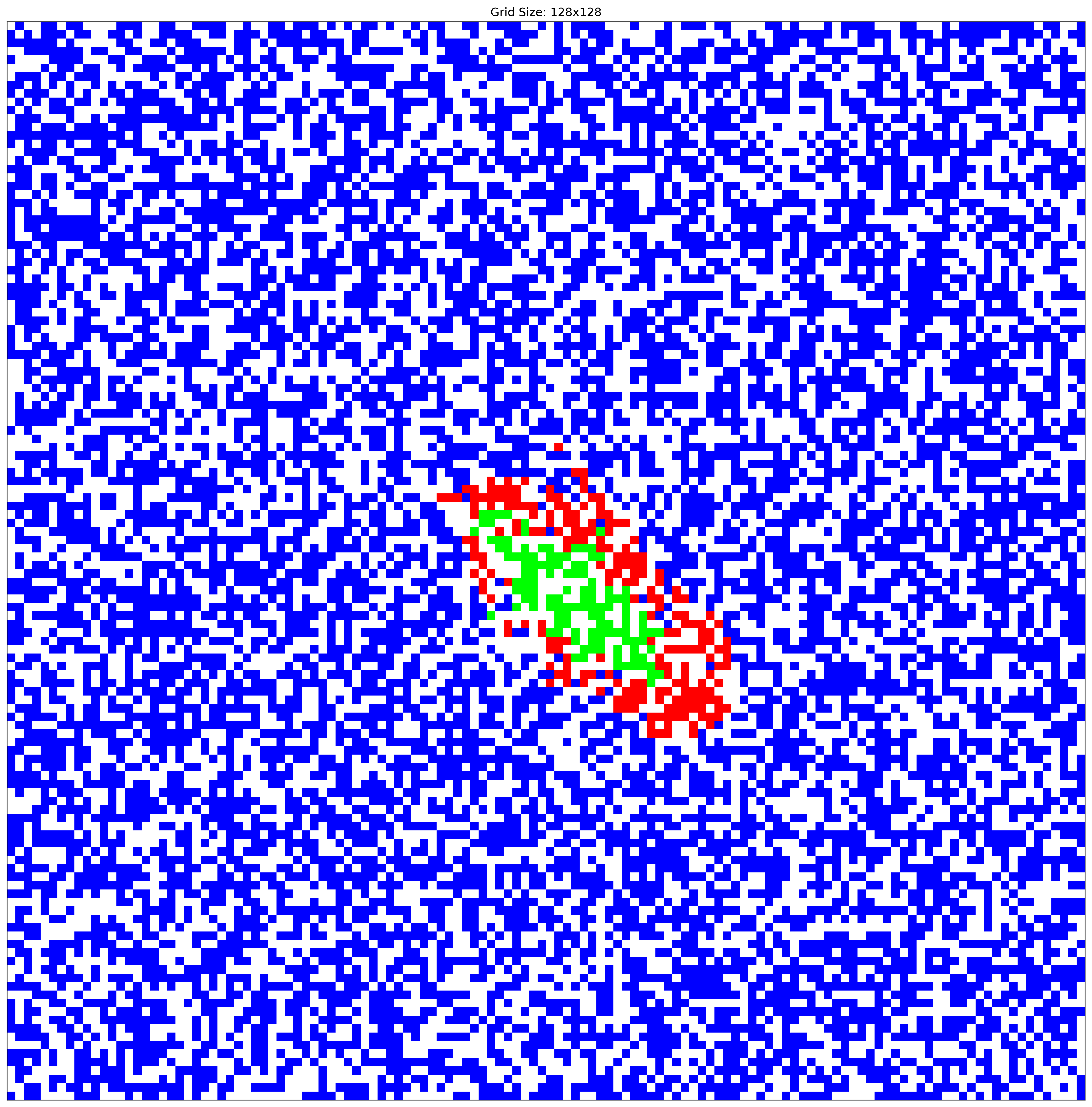}
        \caption{Step = 66}
        \label{fig:step66}
    \end{minipage}%
    \begin{minipage}{0.30\textwidth}
        \centering
        \includegraphics[width=\textwidth]{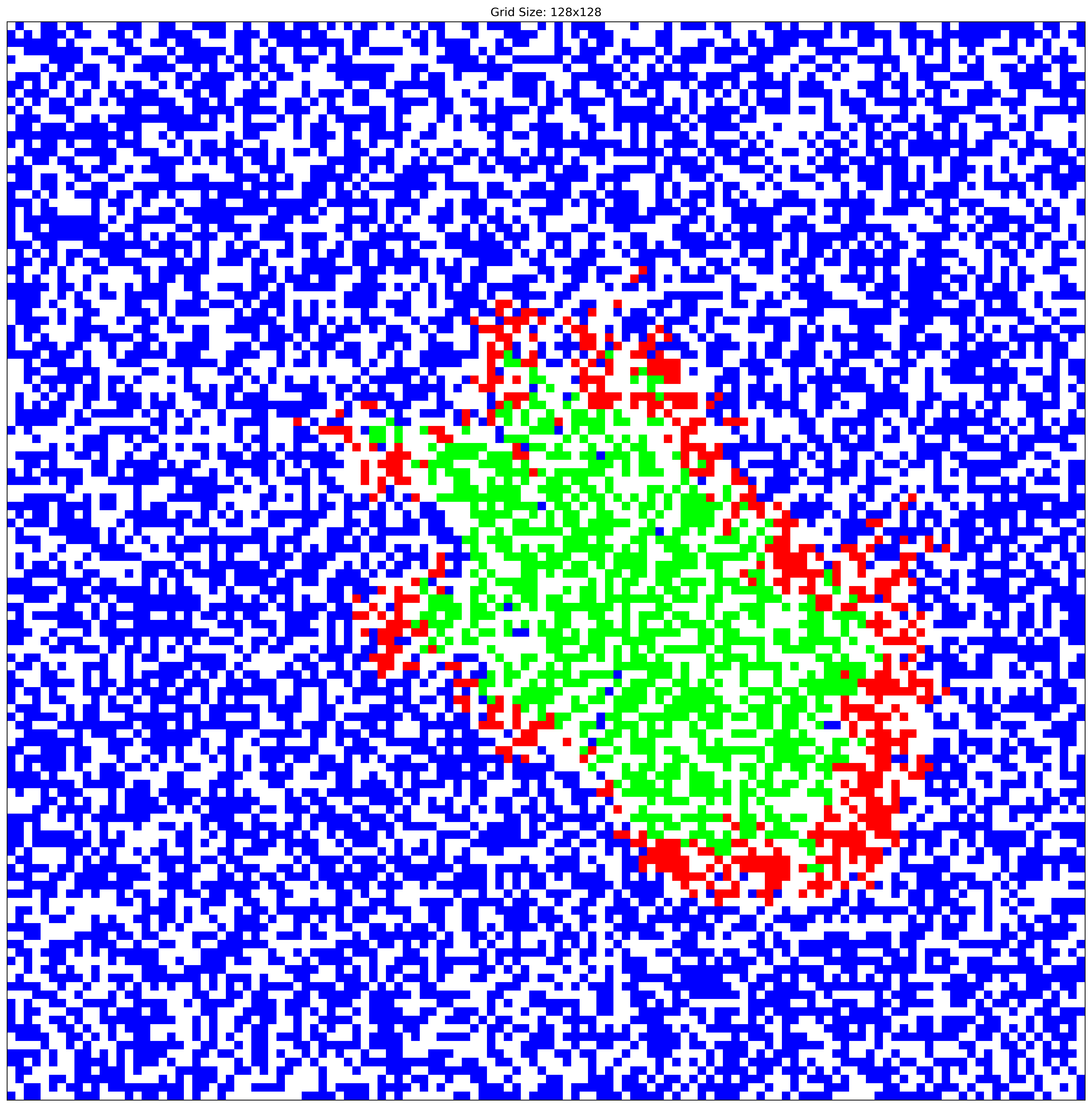}
        \caption{Step = 132}
    \end{minipage}

     \vspace{0.5cm} % Vertical space between rows

    \begin{minipage}{0.30\textwidth}
        \centering
        \includegraphics[width=\textwidth]{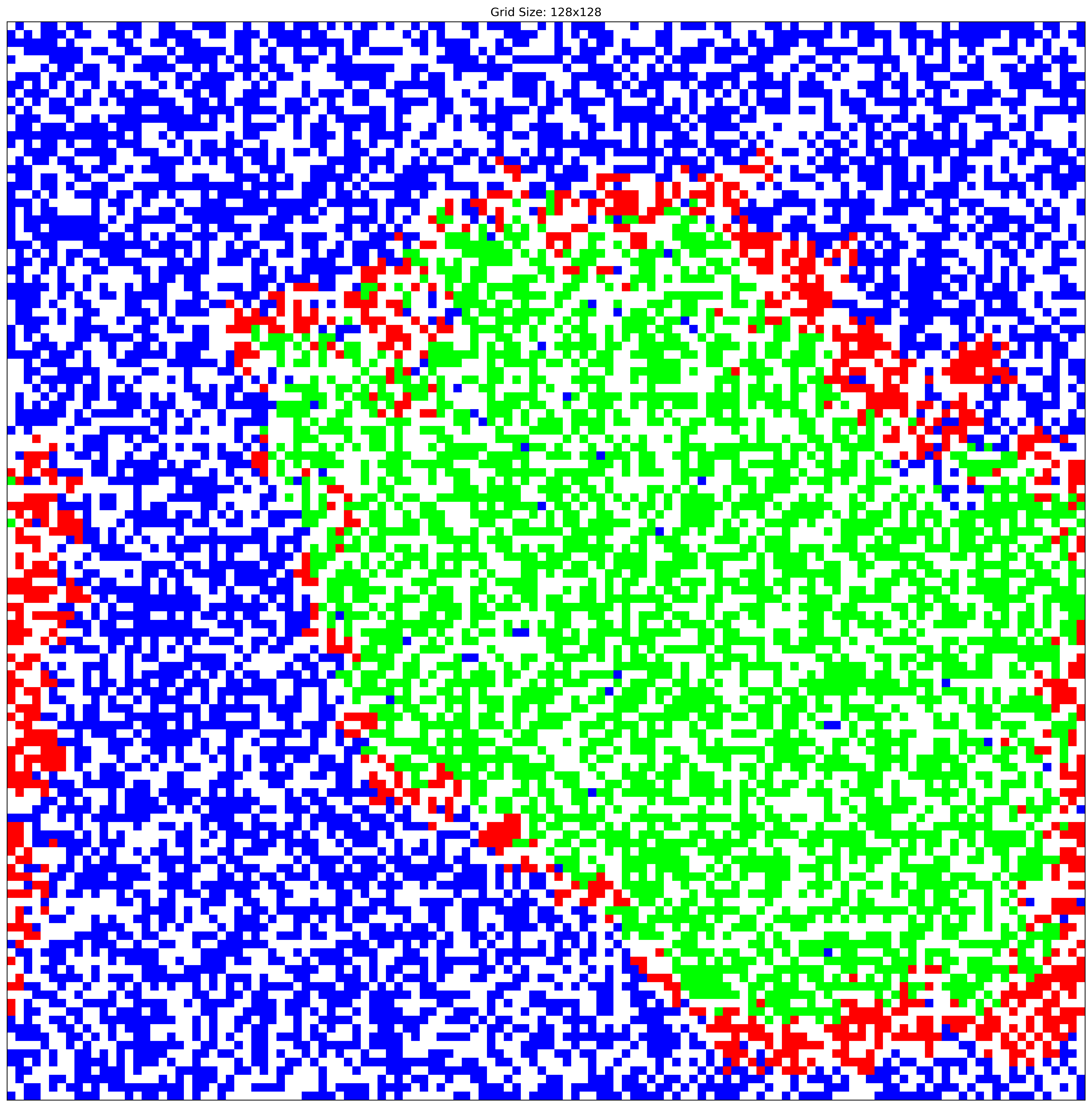}
        \caption{Step = 198}
    \end{minipage}
    \begin{minipage}{0.30\textwidth}
        \centering
        \includegraphics[width=\textwidth]{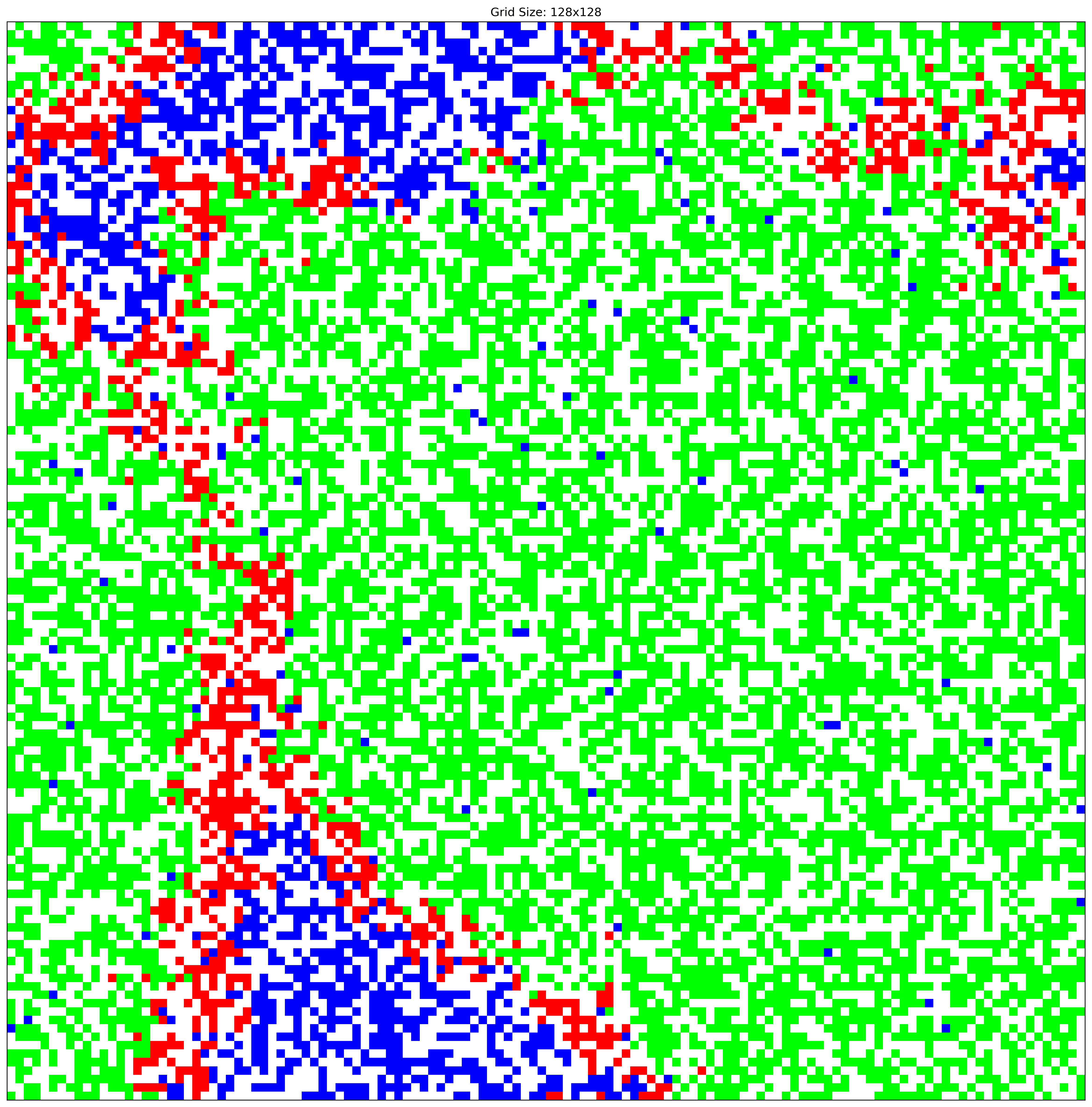}
        \caption{Step = 264}
    \end{minipage}
        \begin{minipage}{0.30\textwidth}
        \centering
        \includegraphics[width=\textwidth]{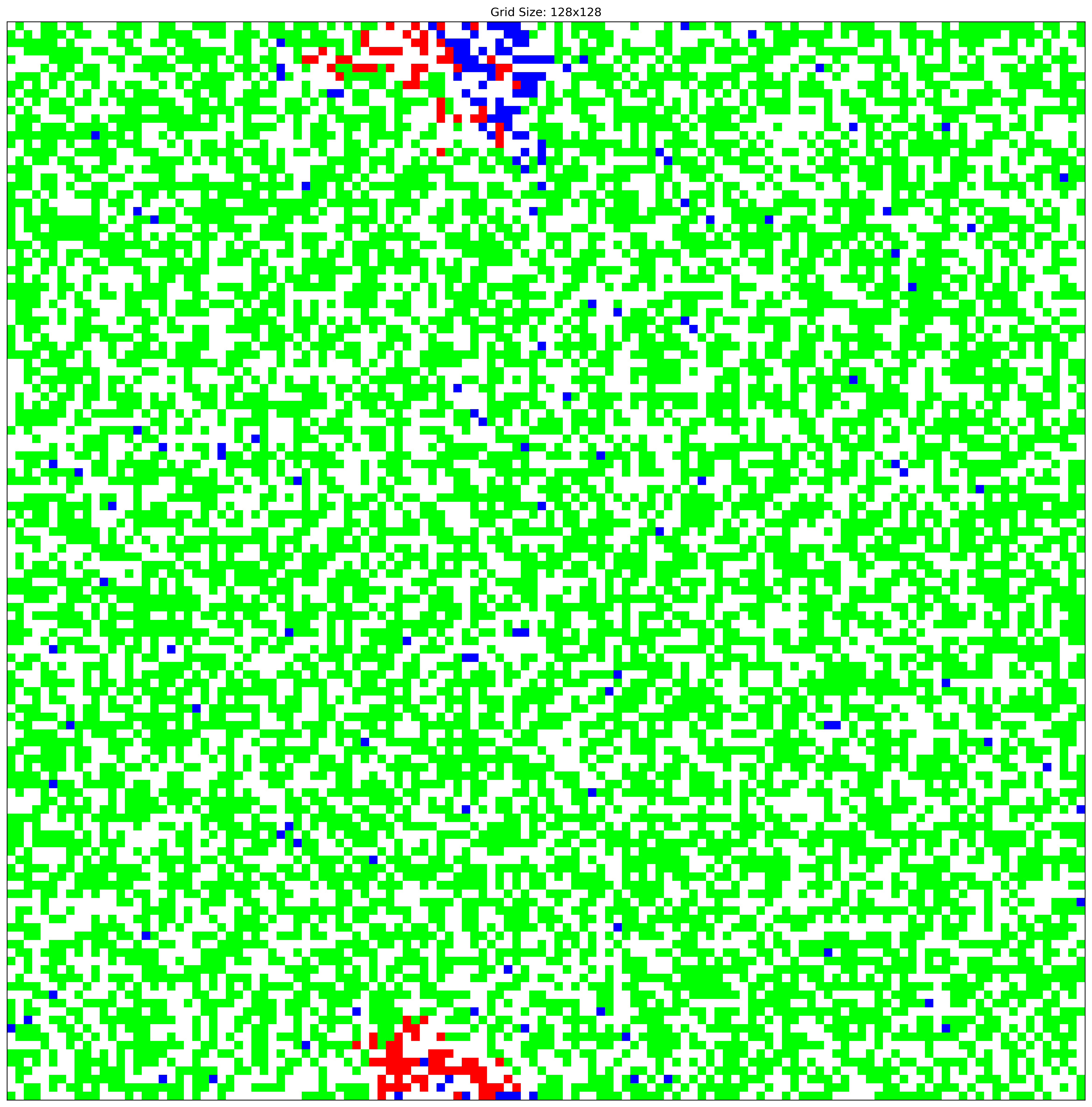}
        \caption{Step = 330}
        \label{fig:step330}
        
         \vspace{0.5cm}
         
    \end{minipage}
        \begin{minipage}{0.30\textwidth}
        \centering
        \includegraphics[width=\textwidth]{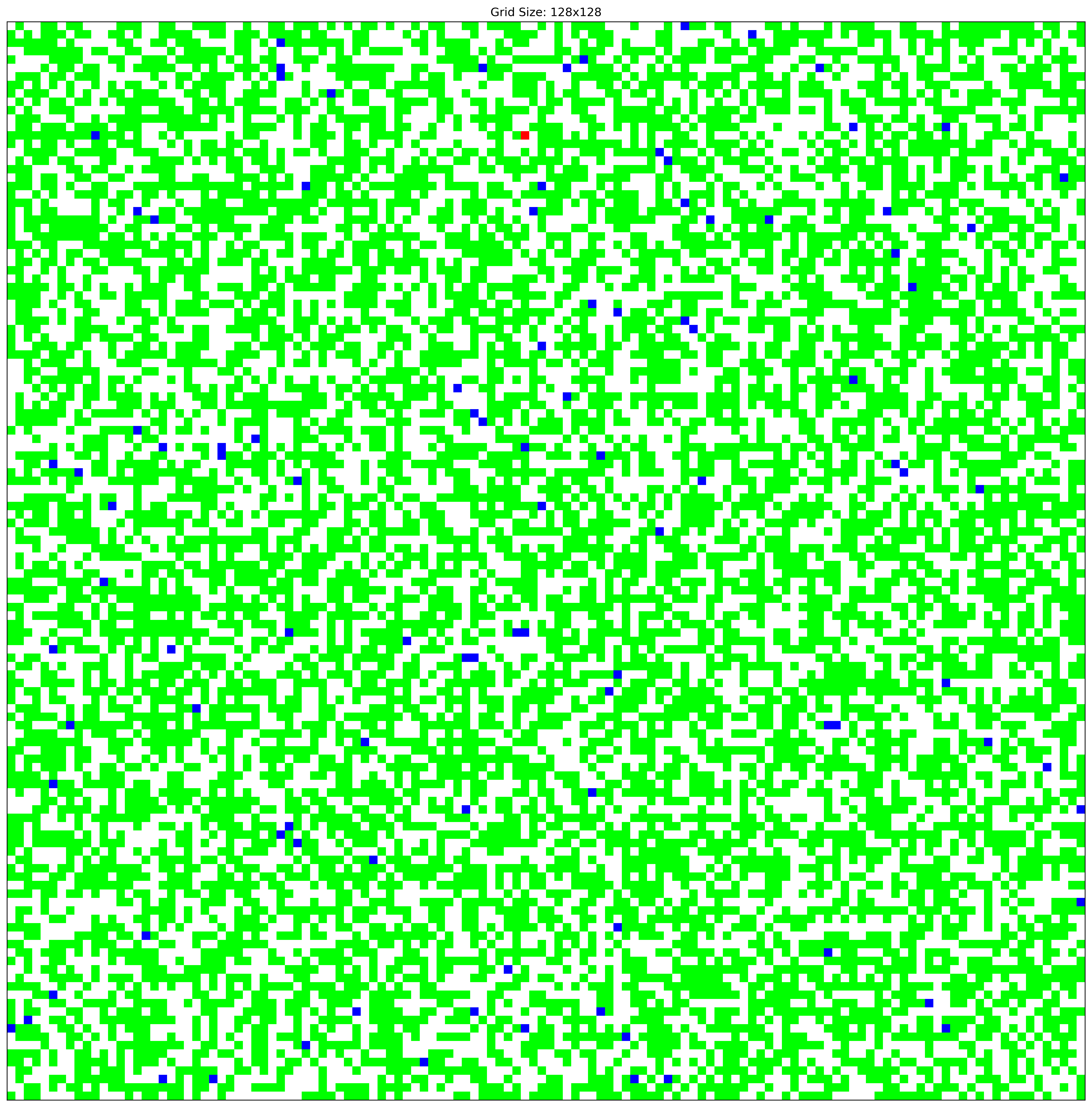}
        \caption{Step=396}
                \label{fig:step396}

    \end{minipage}
            \begin{minipage}{0.30\textwidth}
        \centering
        \includegraphics[width=\textwidth]{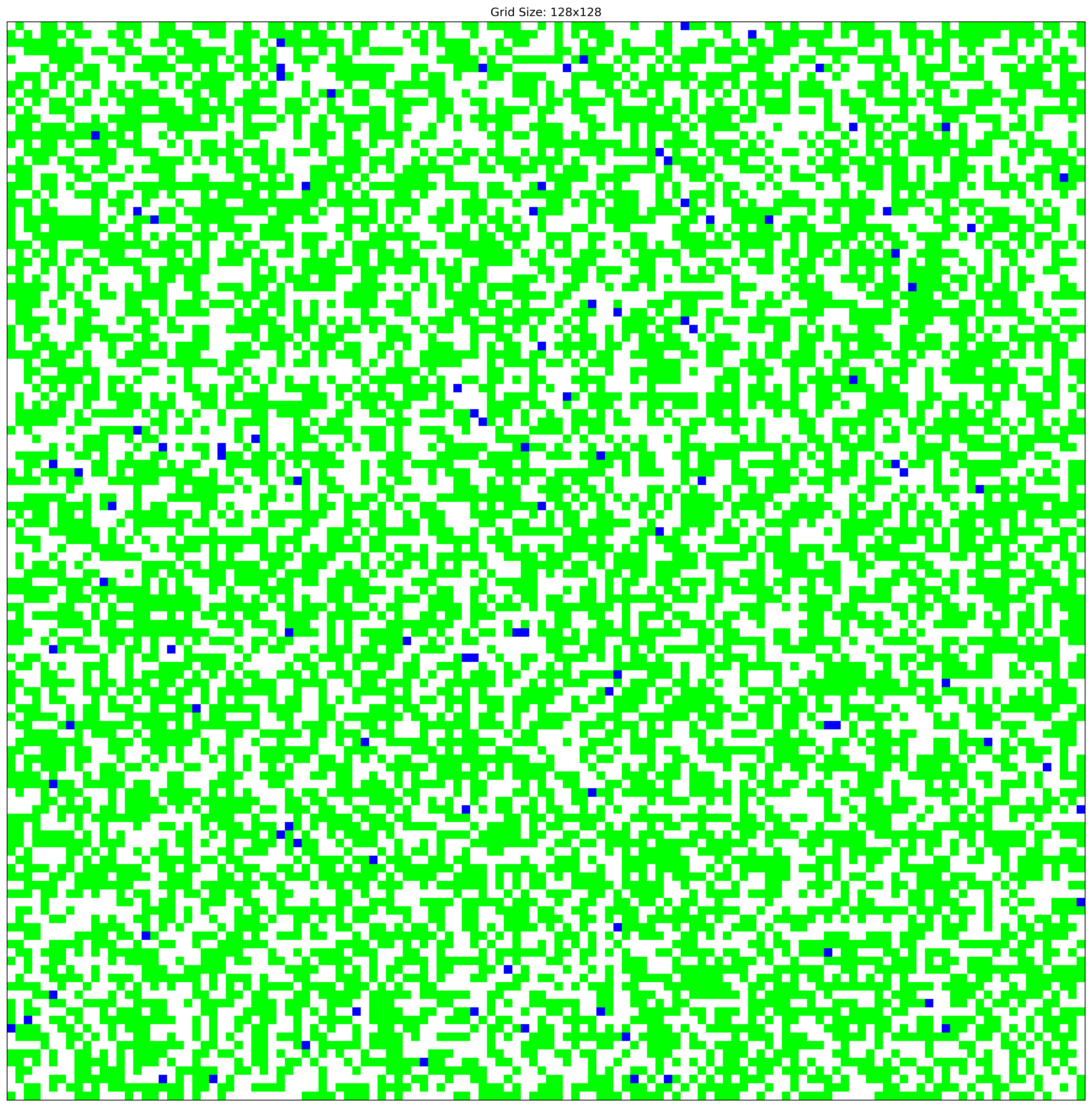}
        \caption{Step=404 (end)}
                \label{fig:step404}

    \end{minipage}
                \begin{minipage}{0.30\textwidth}
        \centering
        \includegraphics[width=\textwidth]{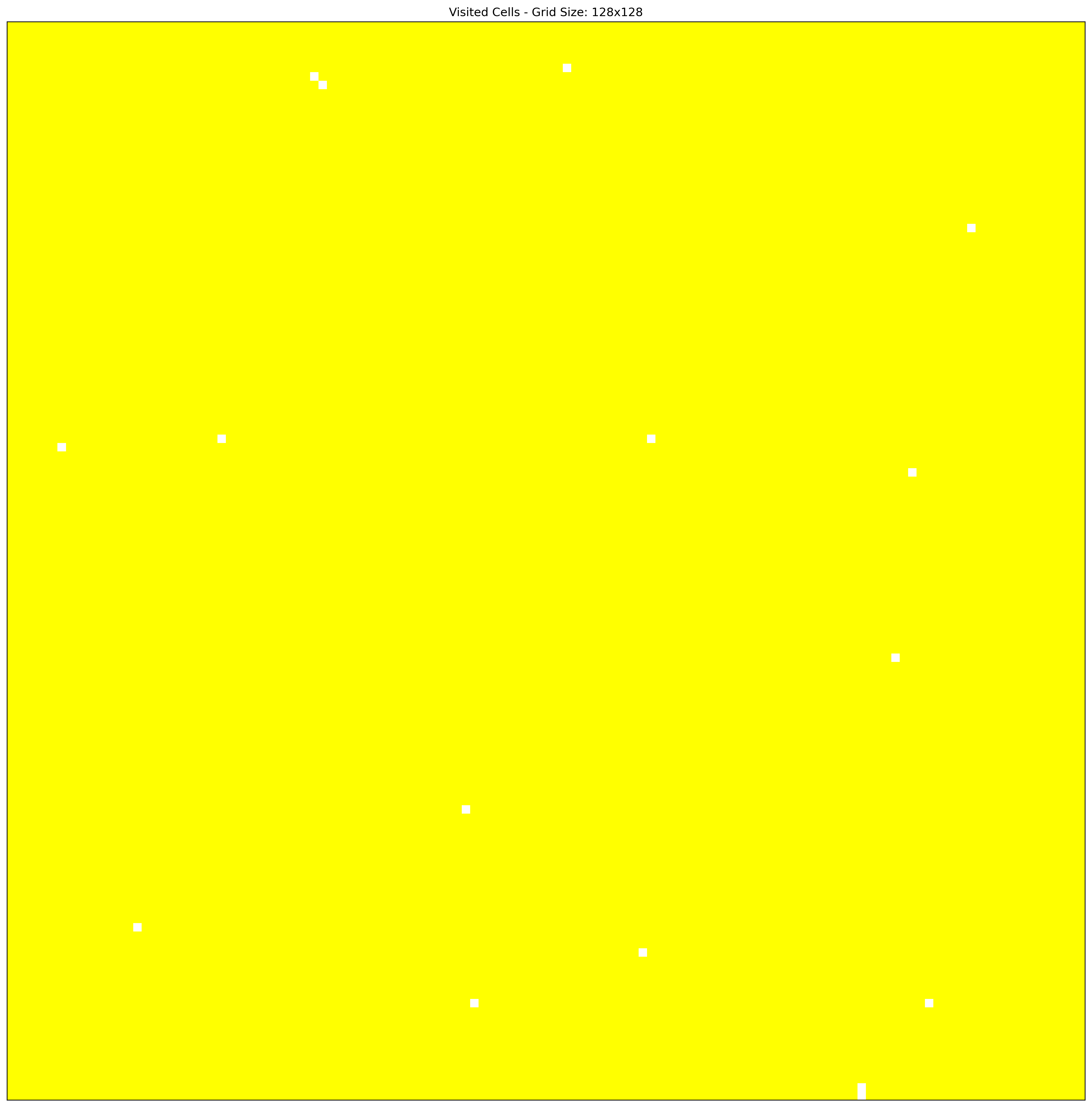}
        \caption{Visited sites}
        \label{fig:visited}
    \end{minipage}

\end{figure}

Figures~\ref{fig:step0}--\ref{fig:visited} illustrate the time evolution of the epidemic under the quantum random walk (QRW) framework. At step $0$ (Figure~\ref{fig:step0}), the lattice contains $N-1$ susceptible agents (blue) and a single infected agent (red). As time progresses (Figures~\ref{fig:step66}--\ref{fig:step330}), the infection propagates anisotropically, forming compact but skewed clusters due to quantum interference effects. By the final stages (Figures~\ref{fig:step396}--\ref{fig:step404}), nearly all susceptible sites have transitioned to recovered, marking the termination of the epidemic. Figure~\ref{fig:visited} shows the complete set of visited sites (yellow), highlighting the spatial extent of the outbreak under QRW dynamics. Compared to the classical random walk case, the spread here is distinctly non-Gaussian and directionally biased, a hallmark of quantum-coherent propagation.
% Another crucial quantity of interest is the \textit{basic reproductive number}, \(R_0\). In epidemiology, \(R_0\) is defined as the expected number of secondary cases directly generated by a single infection in a fully susceptible population. In our grid-based model, this is calculated by marking all agents as susceptible and tracking the number of new infections generated from the initially infected agent.  

% A naive approximation for \(R_0\) is given by
% \[
% R_0 \approx p \, \tau_{\max},
% \]
% where \(p\) is the infection probability. However, this approximation often fails in practice. In the classical random walk case, \cite{chu2021random} proposed the following improved expression:
% \[
% R_0 = \frac{p \, \tau}{1 + K p \ln(\tau / \tau_0)},
% \]
% where \(K\) and \(\tau_0\) are model-dependent constants.  

% We tested the applicability of this formula to the quantum random walk case. However, due to the inherent bias introduced by quantum interference effects, the walker is less likely to return to the origin, where more recovered sites would be concentrated. Thus, intuition suggests that this formula will not hold consistently in the quantum regime. 

\section{Results}
We evaluated the basic reproduction number $R_0$ for various values of the infection probability $p$ and the infection lifetime $\tau_{\max}$. The computed values are summarized in Table~\ref{tab:R0_results}. It is evident that the $R_0$ obtained for the quantum random walk model differs significantly at some $p, \tau$ value from that of the classical case. All simulations were performed on a $64 \times 64$ lattice, corresponding to a population of 4096 agents initially in the susceptible state, with a single infected agent introduced at the origin. For each parameter set, the average value of $R_0$ was estimated over 10,000 independent iterations to ensure statistical reliability.
% \begin{figure}[!ht]
% \centering
% \includegraphics[width=0.8\linewidth]{photos/classical.png}
% \caption{$R_0$ values of classical model from \cite{chu2021random}}
% \end{figure}

\begin{table}[htbp]
  \centering
  
  \begin{tabular}{|c|c|c|c|c|c|}
    \hline
    $\tau$ & $p=1$ & $p=0.5$ & $p=0.25$ & $p=0.125$ & $p=0.0625$  \\
    \hline
    1 & 0.99927 (2)  & 0.4997 (3)  & 0.2498 (3)  & 0.1252 (2)  & 0.0627 (2) \\
    \hline
    2 & 1.8614 (2) & 0.9307 (4)  & 0.4658 (4)  & 0.2339 (3)  & 0.1171 (2) \\
    \hline
    3 & 2.67771 (0)    & 1.37355 (0) & 0.69514 (0)  & 0.34982 (0)  & 0.17546 (0) \\
    \hline
  \end{tabular}
  \caption{$R_0$ values of classical model from \cite{chu2021random}}
  \label{tab:R0C_results}
\end{table}

\begin{table}[htbp]
  \centering
  
  \begin{tabular}{|c|c|c|c|c|c|}
    \hline
    $\tau$ & $p=1$ & $p=0.5$ & $p=0.25$ & $p=0.125$ & $p=0.0625$  \\
    \hline
    1 & 3.6272  & 0.7963  & 0.3057  & 0.1340  & 0.0661 \\
    \hline
    2 & 24.4428 & 1.9397  & 0.5825  & 0.2397  & 0.1022 \\
    \hline
    3 & --    & 6.9934  & 1.2307  & 0.4241  & 0.1867 \\
    \hline
  \end{tabular}
  \caption{$R_0$ values of quantum model for different $\tau$ (rows) and 
  $p$ (columns)}
   \label{tab:R0_results}
\end{table}

% The other important graph that has been created is checking the exponential growth of the size of the cluster which represents the regions visited (marked in yellow) versus the value of \(N\). We ran this simulation on a \(32 \times 32\) dimensional lattice and took the average of 1000 iterations to determine the average cluster size \(<M>\). Using this, we can get the value of the exponent of the equation goverining this curve and then classify the curve according to different universality classes in percolation theory. We can plot this for different lattice sizes and get the exponential curves and check if the value of the exponent generated is consistent across all the curves.
Another important aspect of our study is the growth of the infected cluster, which represents the regions visited by the walker (highlighted in yellow during simulations). To quantify this, we measured the average cluster size $\langle M \rangle$ as a function of the number of steps $N$. The simulations were carried out on a $32 \times 32$ lattice, and the results were averaged over 1000 independent iterations to ensure statistical accuracy. As shown in Fig.~\ref{fig:cluster_growth}, the cluster size $\langle M \rangle$ exhibits exponential growth with increasing $N$.

\begin{figure}[H]
\centering
\includegraphics[width=0.8\linewidth]{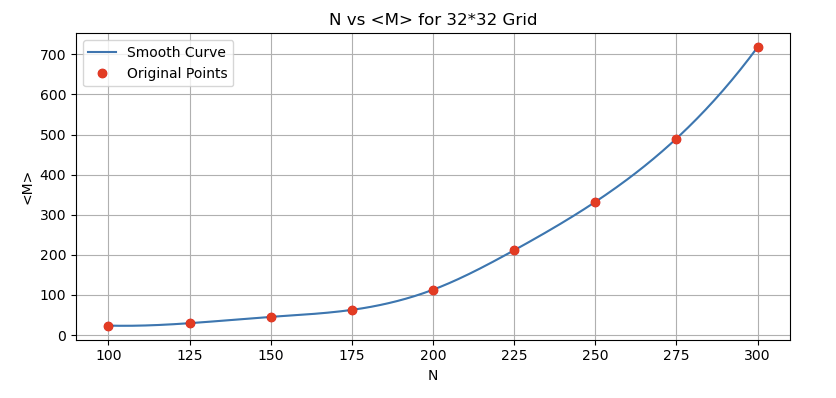}
\caption{\label{fig:result_1}\(<M>\) vs \(N\) curve for $32\times32$ dimensional lattice}
\end{figure}

\section{Discussion}

\subsection{Classical vs. Quantum RW--SIR Model Predictions}

Tables~1 and~2 present the $R_0$ values computed under the classical random walk SIR model (Chu et al.~\cite{chu2021random}) and the quantum random walk SIR model (this work). In the classical random-walk model on a two-dimensional lattice, the basic reproduction number $R_0$ grows roughly linearly with both infection probability $p$ and duration $\tau$ (the ``naïve'' estimate is $R_0 \approx p \cdot \tau$). However, because a two-dimensional walker frequently revisits sites, the true $R_0$ falls below the $p \cdot \tau$ line once paths self-intersect.Indeed, Chu \textit{et al.}~\cite{chu2021random} emphasize that the approximation $R_0 \approx p\tau$ breaks down whenever the trajectory of even a single walker self-intersects.
 Table~1 confirms this: classical $R_0$ is slightly less than $p\tau$ for large $p\tau$ and grows sublinearly. For example, at $p=1$, $\tau=3$ we have $p\tau = 3$ but the classical model gives $R_0 \approx 2.68$.
 In the classical model, the outbreak spreads diffusively: the infection front remains Gaussian and decays rapidly with distance.  

By contrast, the quantum-walk SIR simulations (Table~2) yield dramatically higher $R_0$ at high infectivity. For instance, at $p=1$ and $\tau=2$, the quantum model gives $R_0 \approx 24.44$, versus only $\sim 1.86$ in the classical case. Even at $\tau=1$, $R_0$ jumps from $\sim 0.999$ (classical) to $\sim 3.63$ (quantum) at $p=1$. However, as $p$ decreases, the quantum and classical $R_0$ converge. At low $p$ (e.g.\ $p=0.0625$) the quantum $R_0$ ($\approx 0.0661$ at $\tau=1$) closely matches the classical value ($\approx 0.0627$) for the same parameters.  

In other words, by tuning $p$ the quantum model interpolates between super-diffusive (ballistic) and ordinary diffusive spread. This mirrors the well-known behavior of quantum walks: a coherent (high-$p$) quantum walk spreads quadratically faster than a classical walk, whereas introducing randomness (low $p$) makes it behave diffusively~\cite{kempe2003quantum,venegas2012quantum}. In fact, quantum walks propagate over distance proportional to $t$ (ballistic) versus $\sqrt{t}$ for a classical walk, giving a much broader spatial spread. Thus the quantum RW--SIR can produce very high $R_0$ by effectively ``ballistic'' spread at $p=1$, yet recover the Gaussian-like, lower-$R_0$ classical regime at small $p$. This parameter tunability allows the quantum model to span both extremes: explosive outbreak growth at high $p$ (super-diffusive regime) and standard diffusive spread at low $p$.  

\subsection{Comparison of Quantum Model Data (Table~2) to Real-World $R_0$ Values}

Real-world pathogens typically exhibit $R_0$ values in the range of a few. For example, the original SARS-CoV-2 (Wuhan strain) is estimated at $R_0 \approx 2.2$--$2.8$~\cite{li2020early,liu2020reproductive}, with many studies clustering around $2$--$3$~\cite{park2020estimating}. The Delta variant was far more transmissible, with reported $R_0$ values spanning roughly $3$--$8$~\cite{liu2022delta}, and one review noting Delta’s $R_0$ ``between 3.2 and 8''~\cite{campbell2021increased}. Omicron appears even higher: early data suggest Omicron’s transmissibility is about $3.2\times$ that of Delta~\cite{liu2022omicron}, implying an $R_0$ well into the double digits if completely unchecked. For comparison, the 2003 SARS-CoV outbreak had $R_0 \sim 2$--$3$~\cite{dye2003modeling}, while seasonal influenza viruses are much lower, typically $R_0 \sim 1.0$--$1.5$~\cite{biggerstaff2014estimates}. In short, real outbreaks range from barely supercritical ($R_0 \sim 1$--$2$) to highly explosive ($R_0 \sim 5$--$10$), but rarely exceed $10$.  

The quantum random-walk model produces a much wider range of $R_0$ values than the classical model, spanning from very low ($\approx 0.06$) to extremely high ($\gg 10$). This range overlaps the $R_0$ of many real diseases. The quantum RW--SIR model can reproduce much of this spectrum by adjusting $p,\tau$. For moderate parameters it yields realistic values: e.g.\ with $p \approx 0.5$, $\tau=2$ it gives $R_0 \approx 1.94$, in line with the original COVID range. With $p \approx 0.5$, $\tau=3$ one finds $R_0 \approx 6.99$, comparable to the upper end of Delta/Omicron estimates. By contrast, the purely classical model cannot reach those values (its maximum is $\sim 2.68$ at $\tau=3, p=1$). For example, Measles, one of the most contagious human viruses, has $R_0 \approx 12$--$18$~\cite{measlesR0}, and varicella (chickenpox) $R_0 \approx 10$--$12$~\cite{varicellaR0}; the quantum model exceeds these values under high-$p$, moderate-$\tau$ conditions whereas these values cannot be attained by classical random walk model. At the other extreme, low $p$ yields $R_0 < 1$, mimicking diseases that quickly fizzle out (well below seasonal flu). Only the very high-$p$ quantum regimes produce $R_0$ values beyond empirical experience (e.g.\ $24.4$ at $\tau=2, p=1$ far exceeds any known human virus).  

Crucially, the quantum model’s flexibility also addresses the \emph{shape} of spread. Classical random walks yield Gaussian (diffusive) infection fronts, which do not reflect phenomena like superspreading or rapid cluster seeding. In reality, outbreaks often exhibit heavy-tailed, non-Gaussian patterns. For instance, COVID-19’s early growth was driven by superspreading events (e.g.\ a cruise ship, a religious gathering) that seeded multiple regions in sudden bursts~\cite{endo2020estimating}. Classical diffusion underestimates such leaps. Quantum walks, by contrast, are intrinsically non-Gaussian: their probability distribution is nearly flat over a wide region, unlike the exponentially decaying tails of a Gaussian~\cite{kempe2003quantum}. In effect, a quantum-walk SIR model can mimic ``fat-tailed'' spreading, where rare long-distance jumps (analogous to superspreaders or rapid travel) play a large role. Because $p$ tunes the coherence of the walk, the model can smoothly transition from explosive, non-Gaussian outbreaks to ordinary diffusion.  

In summary, the quantum RW--SIR framework can bridge diffusive and super-diffusive epidemics. Its high-$p$ regime produces the extraordinarily large $R_0$ and flat spread of a ballistic process, while low-$p$ yields the modest, Gaussian-like $R_0$ of classical diffusion. This tunability allows it to match the range of modern outbreak dynamics---from milder epidemics (seasonal flu) to superspreading-driven pandemics---in a way that simple classical random walks cannot~\cite{kempe2003quantum,venegas2012quantum}.

\section{Conclusion}
The quantum random-walk model highlights a compelling contrast with classical epidemic models, suggesting that interference and ballistic propagation may provide a framework for modeling faster disease spread. Overall, the quantum-walk model’s $R_0$ versus $(p,\tau)$ behavior aligns better with observed epidemiology. Its $R_0$ values cover the full spectrum of real pathogens --- from low (e.g.\ influenza) to very high (e.g.\ measles). The quantum RW--SIR framework can bridge diffusive and super-diffusive epidemics. Its high-$p$ regime produces the extraordinarily large $R_0$ and flat spread of a ballistic process, while low-$p$ yields the modest, Gaussian-like $R_0$ of classical diffusion. This tunability allows it to match the range of modern outbreak dynamics. By contrast, the classical random-walk $R_0$ remains modest and grows smoothly, failing to capture these threshold-driven dynamics.

\section{Future Work}
\label{sec:future}

The results presented here point to several natural next steps that will both solidify and extend our findings. First, we will undertake a focused percolation analysis: by sweeping the infection probability \(p\) for multiple lattice sizes and performing finite-size scaling, we aim to estimate a putative critical probability \(p_c\) and extract the associated critical exponents. This will determine whether the QRW-driven outbreaks occupy a universality class distinct from the classical random-walk SIR benchmark. Second, we will pursue semi-analytic approximations for the expected number of distinct sites visited by a QRW of lifetime \(\tau\) and use these to build predictive expressions for reproduction number $R_0$ as function of $\tau,p$; such formulas would complement and help interpret the simulation results. Finally, we will test robustness by varying coin operators and introducing controlled decoherence to interpolate between quantum and classical regimes.

\section*{Code Availability}

The source code used to implement the quantum walk-based epidemiological simulations is openly available at: \\
\href{https://github.com/sayanmanna1/Quantum-Walk-based-Epidemiological-Modeling}{\textcolor{blue}{https://github.com/sayanmanna1/Quantum-Walk-based-Epidemiological-Modeling}}

% \subsection{Key Implications}
% While the quantum walk captures the idea that ``coherent'' spreading can be much faster than random diffusion (and hence can generate large $R_0$), its divergence from epidemiological detail means it cannot reliably reproduce how a pathogen propagates in a population. In particular, the unbounded ballistic spread and interference effects highlight phenomena (wave-like propagation, non-monotonic infection pulses) that do not occur in real outbreaks. The model’s main utility may lie in exploring mathematical limits of spread or designing metaphorical algorithms, but it does not realistically capture the nuanced interplay of biology, behavior, and interventions that determine COVID-19’s $R_0$.  

\bibliographystyle{plain}

\bibliography{references} % The references (bibliography) information are stored in the

\appendix
\section*{Appendix}
\section{ Implementation of 1D Quantum Random Walk on Cyclic Graph}
\label{appendix:1d}
\begin{figure}[H]
\begin{center}
\includegraphics[width=0.5\textwidth]{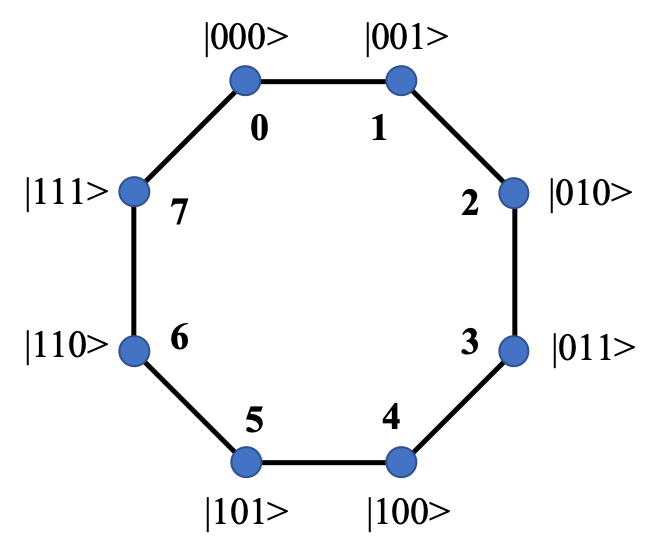}
\end{center}
\caption{8-vertex cycle with position encoded with 3 qubits}
\label{fig:8cycle}
\end{figure}

Here we consider a 1D cycle chain of 8 nodes is represented by the Figure~\ref{fig:8cycle}

As it has 8 vertices, we need 3 bits to represent the position. So, to encode the position, we need $n=log_2{8}=3 $ qubits.
And to choose the direction for moving we need 1 qubit. Therefore total we need $n=4$ qubits to simulate the quantum on this cyclic graph.

A quantum walk on the cycle can be efficiently and straightforwardly implemented with a set of quantum gates consisting of Hadamard gates followed by conditional increment and decrement gates, described below.

\subsection{Mathematical formulation 1D Quantum Walk on cycle}
As shown in the Figure~\ref{fig:8cycle}, node 0 is represented by $|000\rangle$, node 1 by $|001\rangle$, node 2 by $|010\rangle$ and so on.  
\subsubsection{Increment and Decrement Gates}
The Increment Operator denoted by INC increseas the position by 1 step when applied to the current position. Here we considerd a 8 vertex cycle, therefore as an example:\\
\[
\text{INC} \vert 000 \rangle = \vert 001 \rangle
\]
Mathematically, this operator can be expressed as following:
\[
\text{INC} = \vert 000 \rangle \langle 001 \vert + \vert 001 \rangle \langle 010 \vert + \vert 010 \rangle \langle 011 \vert + \vert 011 \rangle \langle 100 \vert +
\]
\[
\vert 100 \rangle \langle 101 \vert + \vert 101 \rangle \langle 110 \vert + \vert 110 \rangle \langle 111 \vert + \vert 111 \rangle \langle 000 \vert
\]

furter we get,
\[
INC = 
\begin{bmatrix}
0 & 0 & 0 & 0 & 0 & 0 & 0 & 1 \\
1 & 0 & 0 & 0 & 0 & 0 & 0 & 0 \\
0 & 1 & 0 & 0 & 0 & 0 & 0 & 0 \\
0 & 0 & 1 & 0 & 0 & 0 & 0 & 0 \\
0 & 0 & 0 & 1 & 0 & 0 & 0 & 0 \\
0 & 0 & 0 & 0 & 1 & 0 & 0 & 0 \\
0 & 0 & 0 & 0 & 0 & 1 & 0 & 0 \\
0 & 0 & 0 & 0 & 0 & 0 & 1 & 0
\end{bmatrix}.
\]

Next the Decrement operator denoted DEC decreases the position by 1 step when applied to the current position.
\[
\text{INC} \vert 000 \rangle = \vert 111 \rangle
\]
Mathematically, this operator can be expressed as following:
\[
\text{DEC}=\vert 001 \rangle \langle 000 \vert + \vert 010 \rangle \langle 001 \vert + \vert 011 \rangle \langle 010 \vert + \vert 100 \rangle \langle 011 \vert +
\]
\[
\vert 101 \rangle \langle 100 \vert + \vert 110 \rangle \langle 101 \vert + \vert 111 \rangle \langle 110 \vert + \vert 000 \rangle \langle 111 \vert
\]

further we get,
\[
DEC = 
\begin{bmatrix}
0 & 1 & 0 & 0 & 0 & 0 & 0 & 0 \\
0 & 0 & 1 & 0 & 0 & 0 & 0 & 0 \\
0 & 0 & 0 & 1 & 0 & 0 & 0 & 0 \\
0 & 0 & 0 & 0 & 1 & 0 & 0 & 0 \\
0 & 0 & 0 & 0 & 0 & 1 & 0 & 0 \\
0 & 0 & 0 & 0 & 0 & 0 & 1 & 0 \\
0 & 0 & 0 & 0 & 0 & 0 & 0 & 1 \\
1 & 0 & 0 & 0 & 0 & 0 & 0 & 0
\end{bmatrix}.
\]
Now to create the conditional translation operator $S$, we define the spin up state by $|\uparrow\rangle$ and spin down state state as $|\downarrow\rangle$ as there are only 2 directions in which the walker can move, where 
\[
|\uparrow\rangle =
\begin{pmatrix}
1 \\
0
\end{pmatrix},
\quad
|\downarrow\rangle =
\begin{pmatrix}
0 \\
1
\end{pmatrix}.
\]

We define $S$ as following:
\[
S = |\uparrow\rangle \langle \uparrow| \otimes \text{INC} + |\downarrow\rangle \langle \downarrow| \otimes \text{DEC}
\]

\[
    = \begin{bmatrix} 1 & 0 \\ 0 & 0 \end{bmatrix} \otimes INC + \begin{bmatrix} 0 & 0 \\ 0 & 1 \end{bmatrix} \otimes  DEC
\]

Further we  get

\[
\scriptsize
S =
\begin{bmatrix}
0 & 0 & 0 & 0 & 0 & 0 & 0 & 1 & 0 & 0 & 0 & 0 & 0 & 0 & 0 & 0 \\
1 & 0 & 0 & 0 & 0 & 0 & 0 & 0 & 0 & 0 & 0 & 0 & 0 & 0 & 0 & 0 \\
0 & 1 & 0 & 0 & 0 & 0 & 0 & 0 & 0 & 0 & 0 & 0 & 0 & 0 & 0 & 0 \\
0 & 0 & 1 & 0 & 0 & 0 & 0 & 0 & 0 & 0 & 0 & 0 & 0 & 0 & 0 & 0 \\
0 & 0 & 0 & 1 & 0 & 0 & 0 & 0 & 0 & 0 & 0 & 0 & 0 & 0 & 0 & 0 \\
0 & 0 & 0 & 0 & 1 & 0 & 0 & 0 & 0 & 0 & 0 & 0 & 0 & 0 & 0 & 0 \\
0 & 0 & 0 & 0 & 0 & 1 & 0 & 0 & 0 & 0 & 0 & 0 & 0 & 0 & 0 & 0 \\
0 & 0 & 0 & 0 & 0 & 0 & 1 & 0 & 0 & 0 & 0 & 0 & 0 & 0 & 0 & 0 \\
0 & 0 & 0 & 0 & 0 & 0 & 0 & 0 & 0 & 1 & 0 & 0 & 0 & 0 & 0 & 0 \\
0 & 0 & 0 & 0 & 0 & 0 & 0 & 0 & 0 & 0 & 1 & 0 & 0 & 0 & 0 & 0 \\
0 & 0 & 0 & 0 & 0 & 0 & 0 & 0 & 0 & 0 & 0 & 1 & 0 & 0 & 0 & 0 \\
0 & 0 & 0 & 0 & 0 & 0 & 0 & 0 & 0 & 0 & 0 & 0 & 1 & 0 & 0 & 0 \\
0 & 0 & 0 & 0 & 0 & 0 & 0 & 0 & 0 & 0 & 0 & 0 & 0 & 1 & 0 & 0 \\
0 & 0 & 0 & 0 & 0 & 0 & 0 & 0 & 0 & 0 & 0 & 0 & 0 & 0 & 1 & 0 \\
0 & 0 & 0 & 0 & 0 & 0 & 0 & 0 & 0 & 0 & 0 & 0 & 0 & 0 & 0 & 1 \\
0 & 0 & 0 & 0 & 0 & 0 & 0 & 0 & 1 & 0 & 0 & 0 & 0 & 0 & 0 & 0
\end{bmatrix}
\]

Now we create the coin operator. Here we are taking a Hadamard coin. To create that we perform tensor product of $H$ and $I$ which is a 8 dimensional identity matrix defined by  
\[
I =
\begin{bmatrix}
1 & 0 & 0 & 0 & 0 & 0 & 0 & 0 \\
0 & 1 & 0 & 0 & 0 & 0 & 0 & 0 \\
0 & 0 & 1 & 0 & 0 & 0 & 0 & 0 \\
0 & 0 & 0 & 1 & 0 & 0 & 0 & 0 \\
0 & 0 & 0 & 0 & 1 & 0 & 0 & 0 \\
0 & 0 & 0 & 0 & 0 & 1 & 0 & 0 \\
0 & 0 & 0 & 0 & 0 & 0 & 1 & 0 \\
0 & 0 & 0 & 0 & 0 & 0 & 0 & 1
\end{bmatrix}.
\]

Therefore, 
\[
C \otimes I = (H \otimes I) = \frac{1}{\sqrt{2}}
\begin{pmatrix}
1 & 1 \\
1 & -1
\end{pmatrix}
\otimes I
\]
Therefore the total evolution operator of the random walker is given by $U$ expressed as following

\[
U = S(H \otimes I) 
\]

\[
\implies U = 
\frac{1}{\sqrt{2}}
\begin{bmatrix}
0 & 0 & 0 & 0 & 0 & 0 & 0 & 1 & 0 & 0 & 0 & 0 & 0 & 0 & 0 & 1 \\
1 & 0 & 0 & 0 & 0 & 0 & 0 & 0 & 1 & 0 & 0 & 0 & 0 & 0 & 0 & 0 \\
0 & 1 & 0 & 0 & 0 & 0 & 0 & 0 & 0 & 1 & 0 & 0 & 0 & 0 & 0 & 0 \\
0 & 0 & 1 & 0 & 0 & 0 & 0 & 0 & 0 & 0 & 1 & 0 & 0 & 0 & 0 & 0 \\
0 & 0 & 0 & 1 & 0 & 0 & 0 & 0 & 0 & 0 & 0 & 1 & 0 & 0 & 0 & 0 \\
0 & 0 & 0 & 0 & 1 & 0 & 0 & 0 & 0 & 0 & 0 & 0 & 1 & 0 & 0 & 0 \\
0 & 0 & 0 & 0 & 0 & 1 & 0 & 0 & 0 & 0 & 0 & 0 & 0 & 1 & 0 & 0 \\
0 & 0 & 0 & 0 & 0 & 0 & 1 & 0 & 0 & 0 & 0 & 0 & 0 & 0 & 1 & 0 \\
0 & 1 & 0 & 0 & 0 & 0 & 0 & 0 & 0 & -1 & 0 & 0 & 0 & 0 & 0 & 0 \\
0 & 0 & 1 & 0 & 0 & 0 & 0 & 0 & 0 & 0 & -1 & 0 & 0 & 0 & 0 & 0 \\
0 & 0 & 0 & 1 & 0 & 0 & 0 & 0 & 0 & 0 & 0 & -1 & 0 & 0 & 0 & 0 \\
0 & 0 & 0 & 0 & 1 & 0 & 0 & 0 & 0 & 0 & 0 & 0 & -1 & 0 & 0 & 0 \\
0 & 0 & 0 & 0 & 0 & 1 & 0 & 0 & 0 & 0 & 0 & 0 & 0 & -1 & 0 & 0 \\
0 & 0 & 0 & 0 & 0 & 0 & 1 & 0 & 0 & 0 & 0 & 0 & 0 & 0 & -1 & 0 \\
0 & 0 & 0 & 0 & 0 & 0 & 0 & 1 & 0 & 0 & 0 & 0 & 0 & 0 & 0 & -1 \\
1 & 0 & 0 & 0 & 0 & 0 & 0 & 0 & -1 & 0 & 0 & 0 & 0 & 0 & 0 & 0 \\
\end{bmatrix}
\]
This $U$ is the evolution operator upon applying repeatedly on a initial state like $|\uparrow\rangle \otimes |000\rangle$, it will evolve and we will get a probability distribution corresponding to the vertices of the cycle after the iterations.

\subsection{Simulation of 1D quantum walk using quantum circuit}
% \begin{figure}[h!]
%     \centering
%     \[
%     \Qcircuit @C=1em @R=1em {
%         % Node group curly brace
%         \lstick{\text{}} & \qw & \multigate{2}{\text{INC}} & \multigate{2}{\text{DEC}} & \qw \\
%                              & \qw & \ghost{\text{INC}}        & \ghost{\text{DEC}}        & \qw \\
%                              & \qw & \ghost{\text{INC}}        & \ghost{\text{DEC}}        & \qw \\
%         % Subnode
%         \lstick{\text{subnode}} & \gate{H} & \ctrl{-1}          & \ctrlo{-1} \qw              & \qw
%     }
%     \]
%     \begin{tikzpicture}[overlay]
%         % Curly brace coordinates
%         \draw [decorate,decoration={brace,amplitude=10pt,mirror}] 
%             (-2.5,3.2) -- (-2.5,1.6) node[midway,xshift=-1cm] {node};
%     \end{tikzpicture}
%     \caption{Quantum circuit with node and subnode components.}
%     \label{fig:quantum_circuit}
% \end{figure}

As shown in Figure~\ref{fig:8cycle}, it is a 8 dimensional cycle, therefore we take 3 qubits to encode the positions denoted by $qc_0, qc_1, qc_2$  and 1 qubit for coin operator denoted by $qanc$ in Figure~\ref{fig:quantum_circuit_3}. 
The quantum circuit for the increment operator is given below in Figure~\ref{fig:quantum_circuit_1}

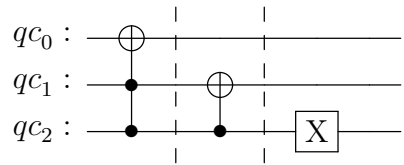
\begin{figure}[H]
    \centering
    \scalebox{1.0}{
\Qcircuit @C=1.0em @R=0.2em @!R { \\
    \nghost{{qc}_{0} :  } & \lstick{{qc}_{0} :  } & \targ \barrier[0em]{2} & \qw & \qw \barrier[0em]{2} & \qw & \qw & \qw & \qw\\
    \nghost{{qc}_{1} :  } & \lstick{{qc}_{1} :  } & \ctrl{-1} & \qw & \targ & \qw & \qw & \qw & \qw\\
    \nghost{{qc}_{2} :  } & \lstick{{qc}_{2} :  } & \ctrl{-1} & \qw & \ctrl{-1} & \qw & \gate{\mathrm{X}} & \qw & \qw\\
\\ }}
    \caption{Quantum circuit to implement the increment operator}
    \label{fig:quantum_circuit_1}
\end{figure}

The quantum circuit for the decrement operator is given below in Figure~\ref{fig:quantum_circuit_2}

\begin{figure}[H]
    \centering
   \scalebox{1.0}{
\Qcircuit @C=1.0em @R=0.2em @!R { \\
    \nghost{{qc}_{0} :  } & \lstick{{qc}_{0} :  } & \qw & \targ & \qw \barrier[0em]{2} & \qw & \qw & \qw & \qw \barrier[0em]{2} & \qw & \qw & \qw & \qw\\
    \nghost{{qc}_{1} :  } & \lstick{{qc}_{1} :  } & \gate{\mathrm{X}} & \ctrl{-1} & \gate{\mathrm{X}} & \qw & \qw & \targ & \qw & \qw & \qw & \qw & \qw\\
    \nghost{{qc}_{2} :  } & \lstick{{qc}_{2} :  } & \gate{\mathrm{X}} & \ctrl{-1} & \gate{\mathrm{X}} & \qw & \gate{\mathrm{X}} & \ctrl{-1} & \gate{\mathrm{X}} & \qw & \gate{\mathrm{X}} & \qw & \qw\\
\\ }}
    \caption{Quantum circuit to implement the decrement operator}
    \label{fig:quantum_circuit_2}
\end{figure}

Now, the whole quantum circuit includes 4th qubit to implement the coin operator, where we use a Hadamard gate to control the direction of walking and increment-decrement operators are being controlled by the 4th qubit $qanc$. $cr$ denotes 3 classical registers where the  first 3 positional qubits will be measured. \cite{douglas2009efficient}

\begin{figure}[H]
    \centering
    \scalebox{1.0}{
\Qcircuit @C=1.0em @R=0.2em @!R { \\
     \nghost{{qc}_{0} :  } & \lstick{{qc}_{0} :  } & \qw \barrier[0em]{3} & \qw & \targ & \qw & \qw \barrier[0em]{3} & \qw & \qw & \targ & \qw & \qw & \qw & \qw & \qw & \qw & \qw\\
     \nghost{{qc}_{1} :  } & \lstick{{qc}_{1} :  } & \qw & \qw & \ctrl{-1} & \targ & \qw & \qw & \gate{\mathrm{X}} & \ctrl{-1} & \gate{\mathrm{X}} & \qw & \targ & \qw & \qw & \qw & \qw\\
     \nghost{{qc}_{2} :  } & \lstick{{qc}_{2} :  } & \qw & \qw & \ctrl{-1} & \ctrl{-1} & \targ & \qw & \gate{\mathrm{X}} & \ctrl{-1} & \gate{\mathrm{X}} & \gate{\mathrm{X}} & \ctrl{-1} & \gate{\mathrm{X}} & \targ & \qw & \qw\\
     \nghost{{qanc} :  } & \lstick{{qanc} :  } & \gate{\mathrm{H}} & \qw & \ctrl{-1} & \ctrl{-1} & \ctrl{-1} & \qw & \gate{\mathrm{X}} & \ctrl{-1} & \qw & \qw & \ctrl{-1} & \qw & \ctrl{-1} & \qw & \qw\\
     \nghost{\mathrm{{cr} :  }} & \lstick{\mathrm{{cr} :  }} & \lstick{/_{_{3}}} \cw & \cw & \cw & \cw & \cw & \cw & \cw & \cw & \cw & \cw & \cw & \cw & \cw & \cw & \cw\\
\\ }}
    \caption{Quantum circuit to implement 1D quantum walk on a 8-vertex cycle}
    \label{fig:quantum_circuit_3}
\end{figure}

Here we show the results of 5 iterations below. We start from the position $|000\rangle$. 
\begin{figure}[H]
    \centering
    \begin{minipage}{0.32\textwidth}
        \centering
        \includegraphics[width=\textwidth]{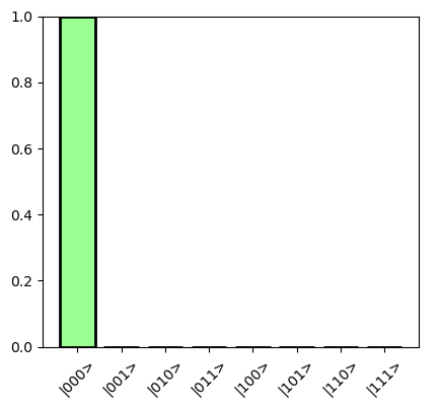}
        \caption{At iteration = 0}
    \end{minipage}%
    \begin{minipage}{0.32\textwidth}
        \centering
        \includegraphics[width=\textwidth]{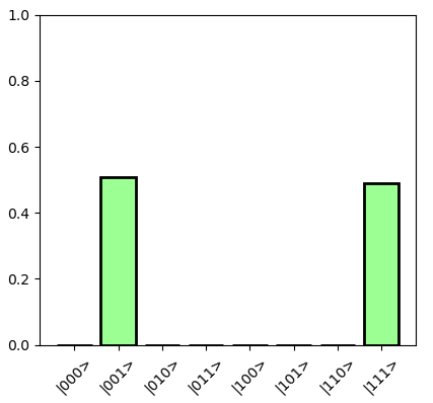}
        \caption{At iteration=1}
    \end{minipage}%
    \begin{minipage}{0.32\textwidth}
        \centering
        \includegraphics[width=\textwidth]{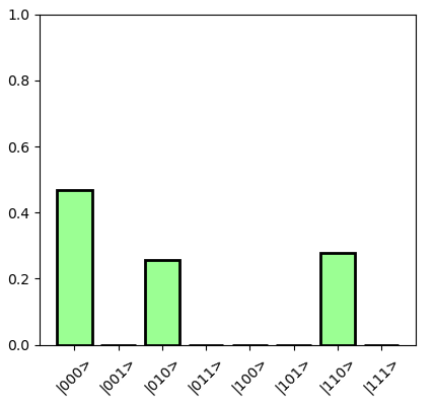}
        \caption{At iteration=2}
    \end{minipage}

    \vspace{0.5cm} % Vertical space between rows

    \begin{minipage}{0.32\textwidth}
        \centering
        \includegraphics[width=\textwidth]{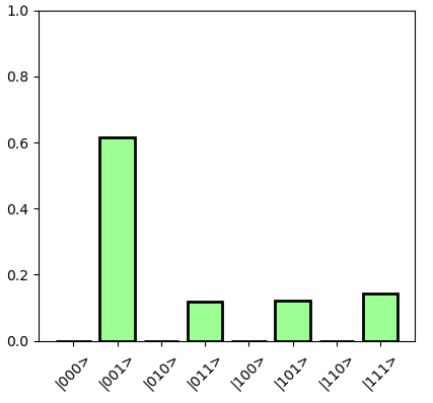}
        \caption{At iteration=3}
    \end{minipage}%
    \begin{minipage}{0.32\textwidth}
        \centering
        \includegraphics[width=\textwidth]{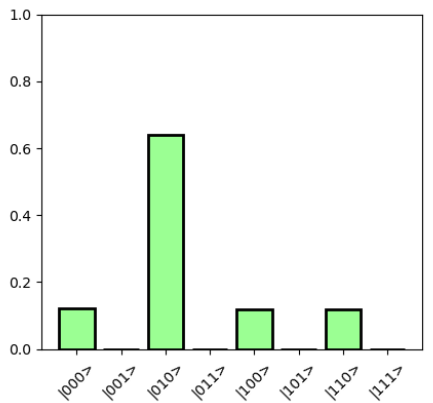}
        \caption{At iteration=4}
    \end{minipage}%
    \begin{minipage}{0.32\textwidth}
        \centering
        \includegraphics[width=\textwidth]{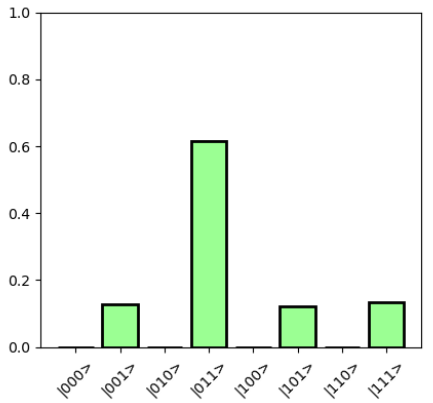}
        \caption{At iteration=5}
    \end{minipage}

    \label{fig:result_1D}
\end{figure}
% \section{Disease spread using Quantum Walk}

% \subsection{1D case}
% \begin{figure}[H]
%     \centering
%     \begin{minipage}{0.32\textwidth}
%         \centering
%         \includegraphics[width=\textwidth]{image/1d_1.png}
%         \caption{At iteration = 0}
%     \end{minipage}%
%     \begin{minipage}{0.32\textwidth}
%         \centering
%         \includegraphics[width=\textwidth]{image/1d_2.png}
%         \caption{At iteration=1}
%     \end{minipage}%
%     \begin{minipage}{0.32\textwidth}
%         \centering
%         \includegraphics[width=\textwidth]{image/1d_3.png}
%         \caption{At iteration=2}
%     \end{minipage}

%     \vspace{0.5cm} % Vertical space between rows

%     \begin{minipage}{0.32\textwidth}
%         \centering
%         \includegraphics[width=\textwidth]{image/1d_4.png}
%         \caption{At iteration=3}
%     \end{minipage}%
%     \begin{minipage}{0.30\textwidth}
%         \centering
%         \includegraphics[width=\textwidth]{image/1d_5.png}
%         \caption{At iteration=4}
%     \end{minipage}%

%     \label{fig:disease_result_1D}
% \end{figure}

\section{Implementation of Quantum Random Walk on a 3D hypercube}
\label{appendix:3d}
\begin{figure}[H]
\begin{center}
\includegraphics[width=0.5\textwidth]{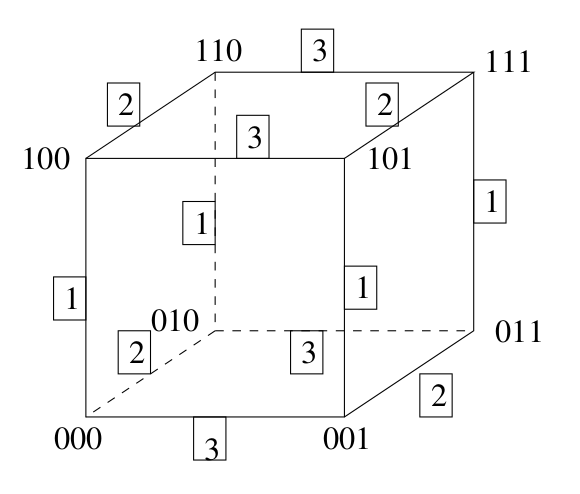}
\end{center}
\caption{The hypercube in d = 3 dimensions. Vertices correspond to 3-bit strings. Edges are labeled by 1, 2, 3 (boxed)
according to which bit needs to be flipped to get from one
vertex of the edge to the other. Adapted from \cite{kempe2003quantum}}
\label{fig:cube}
\end{figure}

We consider a hypercube of dimension 3. There are total 8 vertices and each vertices is denoted 3-bit strings. As shown in the Figure~\ref{fig:cube}, edges are labeled by 1, 2, 3 (boxed). The walker can move to any of the 3 directions labeled by 1,2,3 from a vertex. Now to perform quantum walk, we consider a 3 dimensional coin, which gives 3 different outcomes with equal probability. Let's denote those 3 outcomes by $|1\rangle, |2\rangle, |3\rangle$ corresponding to the directions. Also, the vertices can be encoded by 3 qubits as $|i j k\rangle, i, j, k \in \{1,0\}$. So, if the walker starts from position $|000\rangle$, if we get the outcome $|1\rangle$ from the coin, according to Figure~\ref{fig:cube}, walker will move to $|100\rangle$, now after arriving at $|100\rangle$, if the outcome of the coin comes $|2\rangle$, it will move to $|110\rangle$. Thus the quantum walk will be performed on a hypercube. Here we want to see that how the probability distribution of the position of the walker evolves as it performs quantum walk with 3D coin.

\subsection{Mathematical Formulation}
The coin operator can be generalized by the Discrete Fourier Transform and using this operator, every direction is obtained with
equal probability if we measure the coin space. The matrix for dimension $d$ is give by
\[
DFT = 
\frac{1}{\sqrt{d}}
\begin{bmatrix}
1 & 1 & 1 & \cdots & 1 \\
1 & \omega & \omega^2 & \cdots & \omega^{d-1} \\
1 & \omega^2 & \omega^4 & \cdots & \omega^{2(d-1)} \\
\vdots & \vdots & \vdots & \ddots & \vdots \\
1 & \omega^{d-1} & \omega^{2(d-1)} & \cdots & \omega^{(d-1)(d-1)}
\end{bmatrix},
\]
where \( \omega = e^{2\pi i / d} \) is the primitive \( d \)-th root of unity \cite{kempe2003quantum}
Clearly, the unitary DFT-coin transforms each direction into
an equally weighted superposition of directions such that
after measurement each of them is equally likely to be
obtained (with probability 1/d).\\
For our case, as we are performing quantum walk in a 3 dimensional hypercube with 3 dimensional balanced coin operator, we take $d=3$, therefore the DFT matrix will be given by

\[
DFT_{3d} = 
\frac{1}{\sqrt{3}}
\begin{bmatrix}
1 & 1 & 1 \\
1 & \omega & \omega^2 \\
1 & \omega^2 & \omega^4
\end{bmatrix},
\]

where \( \omega = e^{2\pi i / 3} \) is the primitive 3rd root of unity, satisfying:
\[
\omega^3 = 1 \quad \text{and} \quad \omega^k = e^{2\pi i k / 3}.
\]

Explicitly, this matrix becomes:

\[
DFT_{3d} = 
\frac{1}{\sqrt{3}}
\begin{bmatrix}
1 & 1 & 1 \\
1 & e^{2\pi i / 3} & e^{4\pi i / 3} \\
1 & e^{4\pi i / 3} & e^{8\pi i / 3}
\end{bmatrix}.
\]

For the movement of the walker, now we will define the Conditional Translation Operator $S$.\\

From the Figure~\ref{fig:cube}, we can observe that in direction 1 i.e., if the coin state is $|1\rangle$, then the walker will move from $|000\rangle$ to $|100\rangle$ and vice versa. Similarly, if the coin state is $|2\rangle$, then it will move from $|000\rangle$ to $|010\rangle$. Same thing will happen with all the other nodes as shown in the Figure~\ref{fig:cube}.\\
Therefore we define $S$ as following:
\[
S = |1\rangle\langle 1| \otimes \Big( 
|100\rangle\langle 000| + |000\rangle\langle 100| + |010\rangle\langle 110| + |110\rangle\langle 010| 
\]
\[
+ |101\rangle\langle 001| + |001\rangle\langle 101| + |011\rangle\langle 111| + |111\rangle\langle 011| 
\Big)
\]
\[
+ \, |2\rangle\langle 2| \otimes \Big( 
|110\rangle\langle 100| + |100\rangle\langle 110| + |101\rangle\langle 111| + |111\rangle\langle 101| 
\]
\[
+ |000\rangle\langle 010| + |010\rangle\langle 000| + |001\rangle\langle 011| + |011\rangle\langle 001| 
\Big)
\]
\[
+ \, |3\rangle\langle 3| \otimes \Big( 
|001\rangle\langle 000| + |000\rangle\langle 001| + |011\rangle\langle 010| + |010\rangle\langle 011| 
\]
\[
+ |101\rangle\langle 100| + |100\rangle\langle 101| + |110\rangle\langle 111| + |111\rangle\langle 110| 
\Big).
\]
As an example the initial state of the walker can be $|1\rangle \otimes |000\rangle$, then

\[
S |1\rangle \otimes |000\rangle = |1\rangle \otimes |100\rangle
\]
As here also we are encoding position with 3 qubits, therefore the evolution operator of the quantum walker denoted by $U$ in 3D is given by:
\[
U = S \big( \text{DFT}_{3d} \otimes I \big),
\]

where \( \text{DFT}_{3d} \) represents the 3-dimensional Discrete Fourier Transform matrix, and \( I \) is the $8 \times 8$ dimensional identity operator.

Upon applying $U$ on the walker's state repeatedly will result a probability distribution different than classical random walk.

\subsection{Simulation of 3D quantum walk using quantum circuit}
Here we take 3 qubits to encode the position.

 % As the coin operator $DFT_{3d}$ is a $3 \times 3$ matrix, it can't be simulate in quantum circuit. Instead to simulate coin operator we create a equal superposition of 3 basis states $|00\rangle, |10\rangle, |11\rangle$ with only 2 qubits. Here the state $|00\rangle$ represent the direction 1, state $|10\rangle$ represent the direction 2, state $|11\rangle$ represent the direction 3. Let's denote this unitary operator by $H_3$ which creates the equal superposition of $|00\rangle, |10\rangle, |11\rangle$ So, we need total $n=5$ qubits to simulate this quantum walk.\\
First we describe the quantum circuits to change the position of the walker. There are 3 quantum circuits for moving in 3 directions namely 1,2,3 in Figure~\ref{fig:cube}.
The quantum circuits to move along the direction 1, 2, 3 are given below:
\begin{figure}[H]
    \centering
    \begin{minipage}{0.32\textwidth}
        \centering
        \mbox{
            \Qcircuit @C=1.0em @R=0.2em @!R { 
                & \nghost{{q}_{0} :  } & \lstick{{q}_{0} :  } & \gate{\mathrm{X}} & \qw & \qw \\
                & \nghost{{q}_{1} :  } & \lstick{{q}_{1} :  } & \qw & \qw & \qw \\
                & \nghost{{q}_{2} :  } & \lstick{{q}_{2} :  } & \qw & \qw & \qw \\
            }
        }
        \caption{Direction 1 }
        \label{fig:quantum_circuit_4}
    \end{minipage}
    \hfill
    \begin{minipage}{0.32\textwidth}
        \centering
        \mbox{
            \Qcircuit @C=1.0em @R=0.2em @!R { 
                & \nghost{{q}_{0} :  } & \lstick{{q}_{0} :  } & \qw & \qw & \qw \\
                & \nghost{{q}_{1} :  } & \lstick{{q}_{1} :  } & \gate{\mathrm{X}} & \qw & \qw \\
                & \nghost{{q}_{2} :  } & \lstick{{q}_{2} :  } & \qw & \qw & \qw \\
            }
        }
        \caption{Direction 2}
        \label{fig:quantum_circuit_5}
    \end{minipage}
    \hfill
    \begin{minipage}{0.32\textwidth}
        \centering
        \mbox{
            \Qcircuit @C=1.0em @R=0.2em @!R { 
                & \nghost{{q}_{0} :  } & \lstick{{q}_{0} :  } & \qw & \qw & \qw \\
                & \nghost{{q}_{1} :  } & \lstick{{q}_{1} :  } & \qw & \qw & \qw \\
                & \nghost{{q}_{2} :  } & \lstick{{q}_{2} :  } & \gate{\mathrm{X}} & \qw & \qw \\
            }
        }
        \caption{Direction 3}
        \label{fig:quantum_circuit_6}
    \end{minipage}
\end{figure}

Now we convert each of the quantum circuit shown above into a quantum gate, so that we can use them in the final simulation of quantum circuit. We will call them $move_{1}$, $move_{2}$, $move_{3}$ respectively after converting the quantum circuits of Figure~\ref{fig:quantum_circuit_4}, ~\ref{fig:quantum_circuit_5}, ~\ref{fig:quantum_circuit_6} into quantum gates corresponding to direction 1,2,3 respectively. 

Next, we need the coin operator of dimension $d=3$ because we have total 3 directions. In the above section we have already discussed the mathematical form of the matrix i.e., $DFT_{3d}$. Now this is a $3\times 3$ matrix and we cannot implement it in quantum circuit because to be a quantum gate a matrix has to be $2^n \times 2^n$ dimensional and also unitary. Therefore to implement this  $DFT_{3d}$ matrix, we will make it $4\times 4$ by incorporating 1 row of zeros and 1 column zeros. Further to make it unitary the 16th element has to be 1. Therefore the matrix will look like 
\[
DFTgate_{3d} = 
\frac{1}{\sqrt{3}}
\begin{bmatrix}
1 & 1 & 1 & 0 \\
1 & e^{2\pi i / 3} & e^{4\pi i / 3} & 0 \\
1 & e^{4\pi i / 3} & e^{8\pi i / 3} & 0\\
0 & 0 & 0 & \sqrt{3}
\end{bmatrix}.
\]
Now it has become a 2 qubit unitary gate and it gives super position of 3 basis states as following:
\[
\text{DFTgate}_{3d} \ket{00} = \frac{\ket{00} + \ket{01} + \ket{10}}{\sqrt{3}}
\]
\[
\text{DFTgate}_{3d} \ket{01} = \frac{\ket{00} + e^{\frac{2\pi i}{3}} \ket{01} + e^{\frac{4\pi i}{3}} \ket{10}}{\sqrt{3}}
\]

\[
\text{DFTgate}_{3d} \ket{10} = \frac{\ket{00} + e^{\frac{4\pi i}{3}} \ket{01} + e^{\frac{8\pi i}{3}} \ket{10}}{\sqrt{3}}
\]

Therefore to create this superposition, we start with the initial state $|00\rangle$
 Notice that probability of occurance of each of the states $|00\rangle, |01\rangle, |10\rangle$ which corresponds to direction 1,2,3 is precisely $\frac{1}{3}$. The quantum circuit to implement this gate or create this superposition of 3 basis states is given below:
 \begin{figure}[H]
    \centering
    \mbox{
        \scalebox{0.75}{
            \Qcircuit @C=1.0em @R=0.2em @!R {
                & \nghost{{q}_{0} :  } & \lstick{{q}_{0} :  } & \gate{\mathrm{U_3}\,(\mathrm{2.297,2.048,-1.24})} & \ctrl{1} & \gate{\mathrm{U_3}\,(\mathrm{\frac{\pi}{4},\frac{-\pi}{2},\frac{\pi}{2}})} & \ctrl{1} & \gate{\mathrm{U_3}\,(\mathrm{0.1233,-\pi,\frac{-\pi}{2}})} & \ctrl{1} & \gate{\mathrm{U_3}\,(\mathrm{2.297,1.902,-1.093})} & \qw & \qw \\
                & \nghost{{q}_{1} :  } & \lstick{{q}_{1} :  } & \gate{\mathrm{U_3}\,(\mathrm{0.8446,-1.093,-1.902})} & \targ & \gate{\mathrm{U_3}\,(\mathrm{3.018,\frac{\pi}{2},-\pi})} & \targ & \gate{\mathrm{U_3}\,(\mathrm{\frac{\pi}{2},\frac{-\pi}{2},-\pi})} & \targ & \gate{\mathrm{U_3}\,(\mathrm{2.297,1.902,0.4777})} & \qw & \qw \\
            }
        }
    }
    \caption{Quantum circuit for the 3D coin operator(\(DFTgate_{3d}\))with $\text{global phase:} 3.81869883$  } 
    \label{fig:quantum_circuit_7}
\end{figure}

In Figure~\ref{fig:quantum_circuit_7}, the $U_3$ gate is a single-qubit rotation gate with 3 Euler angles and it is defined as follows:
\[
U_3(\theta, \phi, \lambda) = R_Z(\phi) R_X\left(-\frac{\pi}{2}\right) R_Z(\theta) R_X\left(\frac{\pi}{2}\right) R_Z(\lambda)
\]
Next, we convert this whole circuit of Figure~\ref{fig:quantum_circuit_7} into a single 2 qubit unitary gate and we name it $DFTgate_{3d}$.
We have considered that state $|00\rangle, |01\rangle, |10\rangle$ corresponds to direction 1,2,3 in Figure~\ref{fig:cube}. Total number of qubits required is $n=5$, 3 qubits to encode the positions and 2 qubits for the coin operator. The total quantum circuit for discrete quantum walk on a hypercube~\ref{fig:cube} is as follows for 1 iteration and starting from the position $|000\rangle$
\begin{figure}[H]
    \centering
    \mbox{
        \scalebox{1.0}{
            \Qcircuit @C=1.0em @R=0.2em @!R {
                & \nghost{{q}_{0} :  } & \lstick{{q}_{0} :  } & \qw & \qw & \multigate{2}{\mathrm{move_{1}}}_<<<{0} & \qw & \qw & \multigate{2}{\mathrm{move_{2}}}_<<<{0} & \qw & \multigate{2}{\mathrm{move_{3}}}_<<<{0} & \qw & \qw & \qw \\
                & \nghost{{q}_{1} :  } & \lstick{{q}_{1} :  } & \qw & \qw & \ghost{\mathrm{move_{1}}}_<<<{1} & \qw & \qw & \ghost{\mathrm{move_{2}}}_<<<{1} & \qw & \ghost{\mathrm{move_{3}}}_<<<{1} & \qw & \qw & \qw \\
                & \nghost{{q}_{2} :  } & \lstick{{q}_{2} :  } & \qw & \qw & \ghost{\mathrm{move_{1}}}_<<<{2} & \qw & \qw & \ghost{\mathrm{move_{2}}}_<<<{2} & \qw & \ghost{\mathrm{move_{3}}}_<<<{2} & \qw & \qw & \qw \\
                & \nghost{{q}_{3} :  } & \lstick{{q}_{3} :  } & \multigate{1}{\mathrm{DFTgate_{3d}}}_<<<{0} & \gate{\mathrm{X}} & \ctrl{-1} & \gate{\mathrm{X}} & \qw & \ctrl{-1} & \gate{\mathrm{X}} & \ctrl{-1} & \gate{\mathrm{X}} & \qw & \qw \\
                & \nghost{{q}_{4} :  } & \lstick{{q}_{4} :  } & \ghost{\mathrm{DFTgate_{3d}}}_<<<{1} & \gate{\mathrm{X}} & \ctrl{-1} & \gate{\mathrm{X}} & \gate{\mathrm{X}} & \ctrl{-1} & \gate{\mathrm{X}} & \ctrl{-1} & \qw & \qw & \qw \\
                & \nghost{\mathrm{{c} :  }} & \lstick{\mathrm{{c} :  }} & \lstick{/_{_{3}}} \cw & \cw & \cw & \cw & \cw & \cw & \cw & \cw & \cw & \cw & \cw \\
            }
        }
    }
    
\caption{Quantum circuit implementing discrete quantum walk with 3D coin on a hypercube of $d=3$}
    \label{fig:quantum_circuit_8}
\end{figure}

Here we show the probability distribution of the position of the quantum walker upto 100 iterations below. We start from the position $|000\rangle$ on the cube.
\begin{figure}[H]
    \centering
    \begin{minipage}{0.32\textwidth}
        \centering
        \includegraphics[width=\textwidth]{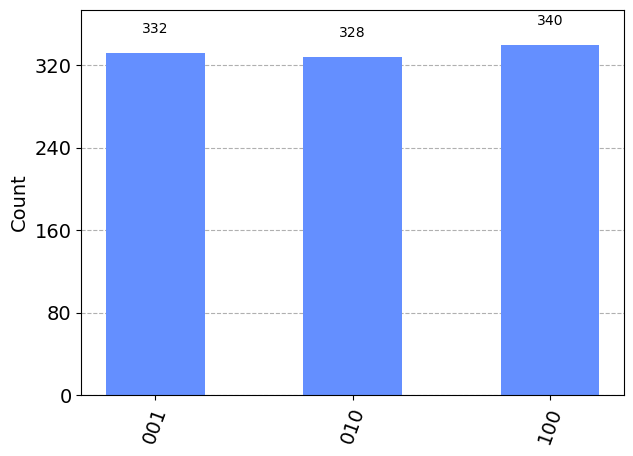}
        \caption{At iteration = 1}
    \end{minipage}%
    \begin{minipage}{0.32\textwidth}
        \centering
        \includegraphics[width=\textwidth]{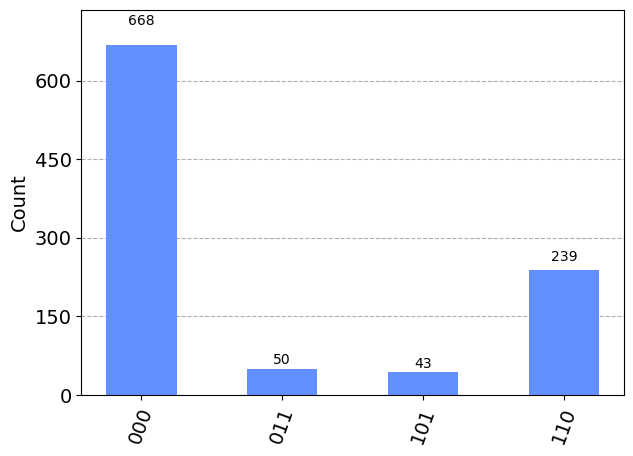}
        \caption{At iteration=20}
    \end{minipage}%
    \begin{minipage}{0.32\textwidth}
        \centering
        \includegraphics[width=\textwidth]{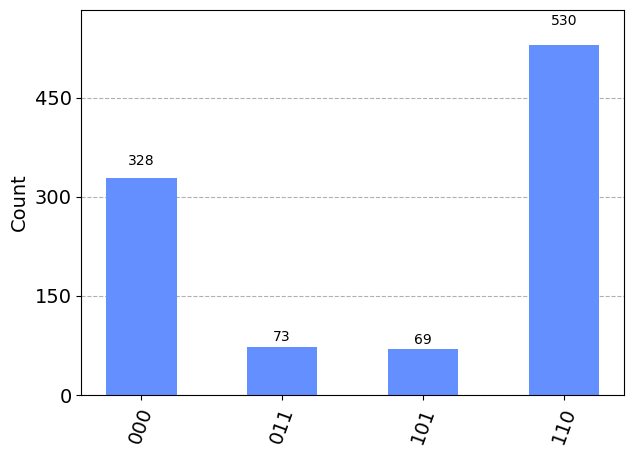}
        \caption{At iteration=40}
    \end{minipage}

    \vspace{0.5cm} % Vertical space between rows

    \begin{minipage}{0.32\textwidth}
        \centering
        \includegraphics[width=\textwidth]{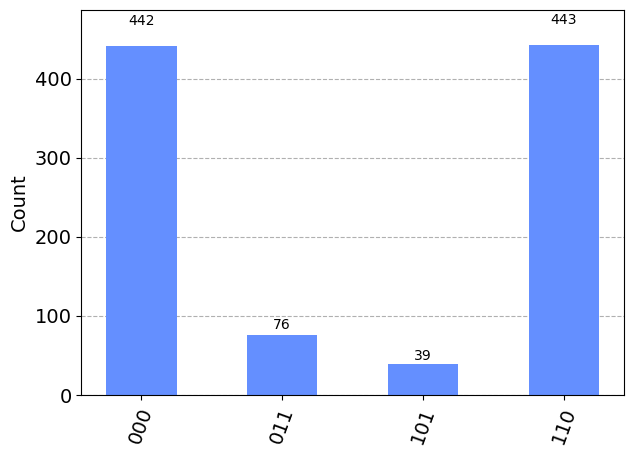}
        \caption{At iteration=60}
    \end{minipage}%
    \begin{minipage}{0.32\textwidth}
        \centering
        \includegraphics[width=\textwidth]{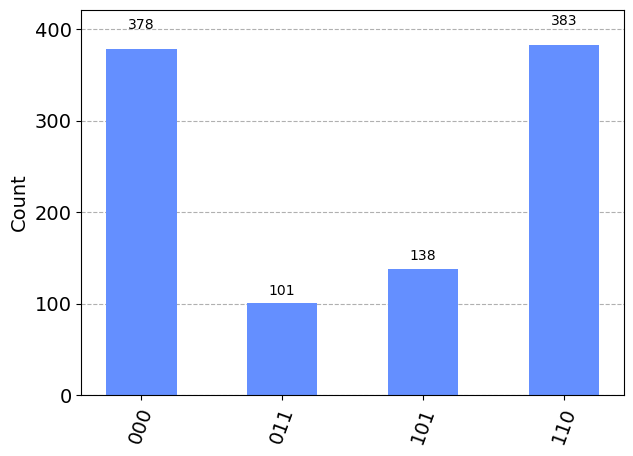}
        \caption{At iteration=80}
    \end{minipage}%
    \begin{minipage}{0.32\textwidth}
        \centering
        \includegraphics[width=\textwidth]{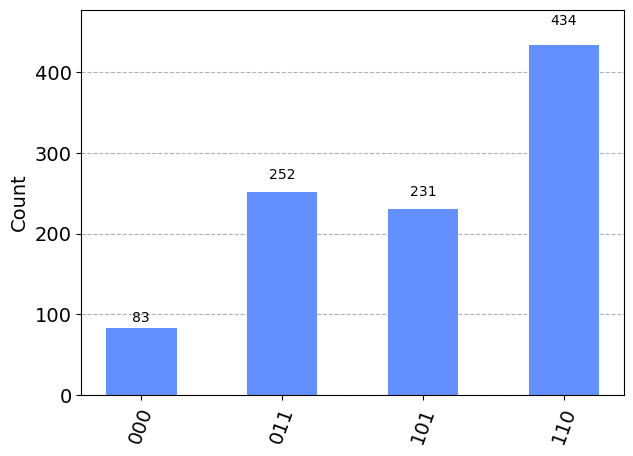}
        \caption{At iteration=100}
    \end{minipage}

    \label{fig:result_3D}
\end{figure}

\end{document}